\documentclass[useAMS,usenatbib]{mn2e}
\usepackage{graphicx,amsmath,txfonts}

\newcommand{\st}{{\it St}}
\title{Vortex instabilities triggered by low-mass planets in pebble-rich, inviscid protoplanetary discs }
\author[A. Pierens, M.-K. Lin, S.~N. Raymond]{A. Pierens $^1$,  M.-K. Lin$^2$, S.~N. Raymond$^1$ \\
$^1$Laboratoire d'Astrophysique de Bordeaux, CNRS and Universit{\'e} de Bordeaux, All{\'e}e Geoffroy St. Hilaire, 33165 Pessac, France  \\
 $^2$  Institute of Astronomy and Astrophysics, Academia Sinica, Taipei 10617, Taiwan}
\date{Released 2012 Xxxxx XX}

\pagerange{\pageref{firstpage}--\pageref{lastpage}} \pubyear{2014}

\def\LaTeX{L\kern-.36em\raise.3ex\hbox{a}\kern-.15em
    T\kern-.1667em\lower.7ex\hbox{E}\kern-.125emX}
\begin{document}
\label{firstpage}
\maketitle
\begin{abstract}
 In the innermost regions of protoplanerary discs, the solid-to-gas ratio can be increased considerably by a number of processes, including photoevaporative  and particle drift.  MHD disc models also suggest the existence of a dead-zone at $R\lesssim 10$ AU, where the regions close to the midplane remain laminar. In this context, we use two-fluid hydrodynamical simulations to study the interaction between a low-mass planet ($\sim 1.7 \;{\rm M_\oplus}$) on a fixed orbit and an inviscid pebble-rich disc with solid-to-gas ratio $\epsilon\ge 0.5$.  For pebbles with Stokes numbers $\st=0.1, 0.5$, multiple dusty vortices  are formed through the Rossby Wave Instability at the planet separatrix. Effects due to gas drag then lead to a strong enhancement in the solid-to-gas ratio, which can increase by a factor of $\sim 10^3$ for marginally coupled particles with $\st=0.5$. As in streaming instabilities, pebble clumps reorganize into filaments that may plausibly collapse to form planetesimals. When the planet is allowed to migrate in a MMSN disc, the vortex instability is delayed due to migration but sets in once inward migration stops due a strong positive pebble torque. Again, particle filaments evolving in a  gap are formed in the disc while the planet undergoes an episode of  outward migration. Our results suggest that vortex instabilities triggered by low-mass planets could play an important role in forming planetesimals in pebble-rich, inviscid discs, and may significantly modify the migration of low-mass planets.  They also imply that planetary dust gaps may not necessarily contain planets if these migrated away.
\end{abstract}

\section{Introduction}

In the standard scenario for planet formation, mm-sized dust grow form from the coagulation of micron-sized grains (Dullemond \& Dominik 2005).  However, further growth  of particles in the mm-size range is  difficult because of the bouncing  (Zsom et al. 2010) and fragmentation (Blum \& Wurm 2008) barriers.  Larger particles are also subject to strong aerodynamic gas drag that causes inward drift on a timescale as short as $\sim 10^2$ yrs for meter-sized bodies at 1 AU (Brauer et al. 2008; Birnstiel et al. 2012).  \\

An emerging picture to bypass these growth barriers is that 100-km sized planetesimals form directly through the streaming instability (Youdin \& Goodman 2005; Johansen et al. 2009; Simon et al. 2016) in which particles with Stokes number (or dimensionless stopping time) $\st\sim 0.001-0.1$  directly concentrate into clumps or filaments under the action of gas drag  and which can subsequently  become gravitationally unstable  to form $\sim100-1000$ km size bodies. These planetesimals can subsequently grow very efficiently by capturing inward drifting pebbles (Johansen \& Lacerda 2010; Lambrechts \& Johansen 2012), namely solids with Stokes number $\st\sim 0.01-1$ that are marginally coupled to the gas,  leading eventually to the formation of  giant planet cores within 1 Myr (Lambrechts \& Johansen 2014). It is expected that pebble accretion should stop once the  mass of the growing embryo has reached the so-called pebble isolation mass, and which corresponds to the critical mass for which  the planet opens a gap and generates a pressure bump in the gas disc outside its orbit that blocks the pebble flux (Morbidelli \& Nesvorny 2012; Bitsch et al 2018; Ataiee et al. 2018). \\

Pebbles also impact planets' orbital evolution.  For a low-mass planet that accretes solids very rapidly, the structure of the gas in its immediate vicinity can be strongly altered by the accretion of solids, resulting in a positive torque that drives outward migration of the planet (Ben\`itez-Llambay et al. 2015).  Pebbles that are scattered by the planet rather than accreted also exert a torque on the planet. Ben\`itez-Llambay \& Pessah (2018)  showed that, when dust feedback is neglected, this occurs because the  interaction with the planet generates strong asymmetries in the dust density distribution. The impact of particle feedback remains to be investigated in more detail as it may play an important role in situations where the local solid-to-gas ratio $\epsilon=\Sigma_p/\Sigma$, where $\Sigma_p$ is the surface density of solids and $\Sigma$ the gas surface density, is locally enhanced. Enhancement in solids is expected in the inner parts of protoplanetary discs as a consequence of the radial drift of solids  or dust settling, but can also arise from photoevaporative (Gorti et al. 2016) or magnetic (Bai et al. 2016) winds.  It can also be due to dust back-reaction, which causes the gas to move outward and be removed from the inner disc.  Kanagawa et al. (2017) showed that the  combined effect of the outward flow of gas and the inward flow of solids can lead to concentrations of solids up to $\epsilon \sim10$, provided that the disc has low viscosity parameter $\alpha \lesssim 10^{-3}$. In the inner parts of the disc, we notice that this appears to a reasonable assumption, since recent simulations that include non-ideal MHD effects find that the disc remains laminar in the region $1<R<10$ AU with a low residual turbulent viscosity ($\alpha < 10^{-4}$; Flock et al. 2017). \\

Here we investigate the consequences of an enhanced solid-to-gas ratio on the interaction between a low-mass planet embedded in an inviscid disc and inward-drifting pebbles with Stokes numbers $\st \ge 0.01$. We pay particular attention to the back-reaction of the pebbles on the gas. We build on the work of Chen \& Lin (2018) who found vortex formation in the co-orbital region for strongly coupled particles $\st \le 10^{-3}$ and dust-to-gas ratios $\epsilon \sim 1$, using an inviscid disc model and following Lin \& Youdin (2017) to model the dusty gas as a single fluid.   Compared to  Chen \& Lin (2018), we therefore consider larger particles that  represent  the main dominant dust species in real protoplanetary discs (Birnstiel et al. 2012). Small dust is rather expected at large orbital distances and at late times in the lifetime of the protoplanetary disc (Lambrechts \& Johansen 2014). Under these conditions, which could be fullfilled for example if the planet migrated outwards or if larger particles are depleted sooner,  the work of Chen \& Lin (2018) may be more relevant than the one presented here.  

We present the results of two-fluid hydrodynamical simulations of marginally coupled particles with $\st \ge 0.01$.   Our simulations assume that one generation of planets has already formed. The significant solid-to-gas ratio that we assume is already above the threshold for planetesimal formation by the streaming instability for certain cases (Carrera et al 2015; Yang et al 2017). While a fraction of that mass should indeed be transformed into macroscopic bodies, most will likely remain in the form of pebbles and dust, and it is that component that we simulate.  It is indeed expected that mm-cm pebbles remain present in the region of giant planet formation over the whole lifetime of the protoplanetary disc (Lambrechts \& Johansen 2014). We show that in dust-rich discs, multiple dusty eddies can form at the separatrix of a low-mass planet as a result of the Rossby  Wave Instability (RWI; Lovelace et al. 1999; Li et al. 2000).  We find that  significant concentrations of solids can be formed, which may promote planetesimal formation. We briefly evaluate the impact of including self-gravity and the effect of planet migration. The formation of vortices  is delayed due to the inward migration of the planet, but finally occurs once a strong positive torque exerted by the pebbles become high enough to stop migration.  \\

The paper is organized as follows. In Sect. 2,  we describe the hydrodynamical model and the initial conditions that are used in the simulations. In Sect. 3, we present the results of our two-fluid numerical simulations. In Sect. 4, we discuss the impact of our results on the planet orbital evolution. We conclude in  Sect. 5.

\section{The hydrodynamic model}
\subsection{Two-fluid model}
\label{sec:steady}

We consider a razor-thin, isothermal protoplanetary disc model and make use of a two-fluid approach to follow the evolution of the gas and solid particles. These are treated as a pressureless fluid, and we expect this approximation to be valid for Stokes numbers $\st\lesssim 0.5$ (Hersant 2009). For the solid component, the equations for the conservation of mass and momentum are given by:
\begin{equation}
\frac{\partial \Sigma_p}{\partial t}+\nabla\cdot(\Sigma_p{\bf V})=0
\end{equation}
\begin{equation}
\frac{\partial {\bf V}}{\partial t}+({\bf V}\cdot \nabla){\bf V}=-{\bf \nabla} \Phi-\frac{{\bf V}-{\bf v}}{\tau}
\label{eq:dust}
\end{equation}
whereas those corresponding to the gas component are given by:
\begin{equation}
\frac{\partial \Sigma}{\partial t}+\nabla\cdot(\Sigma{\bf v})=0
\end{equation}
\begin{equation}
\frac{\partial {\bf v}}{\partial t}+({\bf v}\cdot \nabla){\bf v}=-\frac{\nabla P}{\Sigma}-{\bf \nabla} \Phi-\frac{\Sigma_p}{\Sigma}\frac{{\bf v}-{\bf V}}{\tau}
\end{equation}
where $\Sigma_p$ is the solid surface density, ${\bf V}$ the pebble velocity , $\Sigma$  the gas surface density, $P$ the gas pressure and ${\bf v}$ the gas velocity.  In the previous expressions, $\Phi$ is the gravitational potential which includes the contributions from the star and planet,  the indirect potential that results from the fact that the frame centred onto the central star is not inertial,  and eventually the self-gravity of the gas and solid components. $\tau$ is the friction time that we will parametrize through the Stokes number $\st=\Omega_k \tau$, where $\Omega_k$ is the keplerian angular velocity.  This implies that the strength of the dust-gas coupling is fixed throughout the numerical domain. In a more realistic case of Epstein drag with fixed particle size, the stopping time is proportional to the gas density, so that particles will become loosely coupled if they eventually fall in a gas gap.  Anticipating  the discussion about the results of the simulations later in the paper (see Sect. \ref{sec:results}), we however note that in this work, the gas remains only weakly perturbed because only low-mass planets are considered, with maximum perturbations that are of the order of $15-20\%$  in our reference calculation with $\st=0.1$ (see Sect. \ref{sec:st0v1}). Therefore,  adopting a fixed Stokes number appears to be a reasonable assumption.

 A steady-state solution to the previous equations can be found assuming that both the azimuthal and radial velocities difference $\delta v_\phi$ between the gas and solid particles are much smaller than the bakground velocity that includes the effect of a pressure gradient in the disc  (Nakagawa et al. 1986).  Neglecting all second-order terms in $\delta v_\phi$, one can find that at steady-state,   the radial and azimuthal components of the pebble velocity   must satisfy:
\begin{equation}
V_R=\frac{{\it St}}{(1+\epsilon^2)+{\it St}^2}\Delta v
\label{eq:vrd}
\end{equation}
\begin{equation}
V_\phi=\frac{1+\epsilon}{2((1+\epsilon)^2+\st^2)}\Delta v
\end{equation}
 while those for the gas are given by:
 
 \begin{equation}
 v_R=-\frac{\epsilon \st}{(1+\epsilon^2)+\st^2}\Delta v
 \label{eq:vrg}
\end{equation}
\begin{equation}
v_\phi=\frac{1}{2(1+\epsilon)}\left(1+\frac{\epsilon \st^2}{(1+\epsilon)^2+\st^2}\right)\Delta v
\label{eq:vtg}
\end{equation}

with 
\begin{equation}
\Delta v=\frac{1}{\Sigma\Omega_k}\frac{\partial P}{\partial r}=h^2v_k(2f+s-1)=-\eta v_k
\end{equation}

Here, $\eta=h^2(1-s-2f)$, where $f$ is the flaring index, $s$ the power-law index of the gas surface density profile, and $h$ the disc 
aspect ratio.  At steady-state, the previous expressions for the gas and dust velocities must also satisfy the continuity equations  
$\partial_R(\Sigma_p R V_R)=0$ and $\partial_R(\Sigma R v_R)=0$, which implies that the surface density profile is also completely 
determined, and with a power-law index such that $s+2f+1/2=0$.

\subsection{Numerical method}

Simulations were performed using the GENESIS (De Val-Borro et al. 2006) code which solves
the equations governing the disc evolution on a polar grid $(R,\phi)$ using an advection scheme based on the monotonic  transport algorithm (Van Leer 1977). It uses the FARGO algorithm (Masset 2000) to avoid time step limitation due to the Keplerian velocity at the inner edge of the disc, and was recently extended to follow the evolution of a solid component that is modelled assuming a pressureless fluid.  Momentum exchange between the particles and the gas is handled by employing the semi-analytical scheme presented in Stoyanovskaya et al. (2018). This approach enables considering arbitrary solid concentrations and values for the Stokes number, and  is therefore very well suited for looking for solutions of non-stationary problems.  Tests of the numerical method to handle the momentum transfer between gas and dust are presented in Appendix \ref{sec:test}. The code also includes a module to calculate the self-gravitational potential $\Phi_{sg}$ of the solid and gas components by solving the Poisson equation: 

\begin{equation}
\nabla^2 \Phi_{sg}=4\pi G(\Sigma+\Sigma_p)
\end{equation}

using a Fast Fourier Transform (FFT) method (Binney \& Tremaine 1987; Baruteau \& Masset 2008). In order to take into account the effect of the finite disc thickness, the gravitational potential for the gas is smoothed out using a softening length $r_{s,g}=bR$ with $b=0.7h$ (Muller \& Kley 2012) whereas  the softening length for the dust $r_{s,d}$ is set to $r_{s,d}=0.1r_{s,g}$ (see Sect. \ref{sec:sg} for details). \\
 The computational units that we adopt are such that the unit of mass is the central mass $M_*$, the unit of distance is the initial semi-major axis $a_p$ of the  planet and the gravitational constant is G=1.  To present the results of
simulations we use the planet orbital period as the unit of time.  For most of the simulations presented here, we use $N_R = 848$ radial grid cells uniformly distributed between $R_{in} = 0.4$ and $R_{out} = 1.8$, and $N_\phi=2000$ azimuthal grid cells; but we also considered a few models with double resolution to check the convergency of our results. 

In the limit of tightly dust particles with $\st\rightarrow 0$, the dust+gas disc tends to behave as a single fluid with  reduced sound speed. As a consequence,  the effective scale height ${\tilde H}$ of the dusty gas is also smaller, and is given by (Lin \& Youdin 2017):
\begin{equation}
\tilde H=\frac{H}{\sqrt{1+\epsilon}}, 
\label{eq:heff}
\end{equation}
Our standard resolution is such that $\tilde H$ is resolved by $(21,11)$ cells in the radial and azimuthal directions respectively for our fiducial run with $\epsilon=1$ and $h=0.05$ (see Sect. \ref{sec:init}).  It is important that $\tilde H$ is accurately resolved because we expect the  Rossby Wave Instability, which plays an important role in this work in the development of vortices, to have typical radial wavelenght of the order of $\tilde H$.  

 We also note that in the model where self-gravity is taken into account, we employed a logarithmic radial spacing, as required by the FFT method of Binney \& Tremaine (1987)  \\
To avoid wave reflection at the edges of the computational domain, we employ damping boundary conditions (de Val-Borro et al. 2006),  using wave killing zones for  $R >1.6$ and  $R < 0.5$  where the surface density and velocities for the dust and gas components are relaxed toward their initial values.


\subsection{Initial conditions}
\label{sec:init}

 \indent \par{\bf Gas component--} We consider  locally isothermal disc models that have a constant aspect ratio  $h=0.05$. The initial disc surface density is   $\Sigma=  \Sigma_0(R/R_p)^s$ , where $\Sigma_0=5\times 10^{-4}$ is the initial surface density at the position of the planet  and where the power-law index is set to  $s=-1/2$  so that the constraint $s+2f+1/2=0$ (see Sect. \ref{sec:steady}) is fullfilled. Assuming that the radius $R_p = 1$ in the computational domain corresponds to 5.2 AU, such a value for $\Sigma_0$ would correspond to a disc equivalent to the MMSN and  containing $0.02 M_*$ of gas material inside to 40 AU.  This corresponds to  a value of the Toomre parameter   ${\cal Q}=\kappa c_s/\pi G \Sigma$, where $\kappa$ is the epicyclic frequency and $c_s$ the sound speed, of ${\cal Q}\sim 25$ at the outer edge of the disc.  In simulations where the planet is allowed to migrate, we also considered surface densities that are reduced by factors of 2, 4, 8. Because we focus on inviscid discs, no kinematic viscosity is employed in all the runs presented here. 

\par{\bf Dust component--} The initial dust surface density is such that the initial solid-to-gas ratio (or metallicity) $\epsilon=\Sigma_p/ \Sigma$  is constant throughout the disc, with $\epsilon=0.01, 0.5,$ or $1$. Regarding the Stokes number, our fiducial run has $\st =0.1$ but we also considered  particles with $\st=0.01, 0.5$.\\
For a given set  of input parameters $(\st, \epsilon)$, velocities for both the gas and solid components  can  then be initiated using Eqs. \ref{eq:vrd} to Eqs \ref{eq:vtg}.

\par{\bf Planet--}  We consider a planet with star-to-planet mass ratio of $q=5\times 10^{-6}$ and whose initial semimajor axis is $a_p=1$. For this value of $q$, the standard resolution that is employed is such that the half-width of the horseshoe region $x_s\sim 1.1a_p\sqrt{q/h}\sim 0.01a_p$ (Paardekooper et al. 2010), is resolved by $\sim 7$ grid cells, which is sufficient for the relative error on the corotation torque to be less than $10 \%$ (Masset 2002). Moreover, we have $h/q^{1/3}> 1$ so that the thermal criterion for gap opening (Lin \& Papaloizou 1993) is not satisfied and the planet is not expected to significantly alter the background gas surface density profile. Although for most cases the planet evolves on a fixed circular orbit,  we performed a few runs  in which the planet is allowed to migrate  in order to estimate the impact  of planet migration on our results.   In any case, the planet gravitational potential is smoothed over a smoothing length of $r_{s,p}=0.4 H(a_p)$ which for simplicity is chosen to be the same for both the gas and dust fluids. Since the smoothing length accounts for the vertical stratification of the disc, adopting a smaller smoothing length for the dust than for the gas would be a more appropriate choice.  This is worthwhile to notice because as the instability described in this paper  is excited in the horseshoe region of the planet whose width depends on the adopted smoothing length (Paardekooper et al. 2010), it may be partly controlled by the softening.   \\
 

\section{Results}
\label{sec:results}
\subsection{Impact of the dust feedback}

\begin{figure}
\centering
\includegraphics[width=\columnwidth]{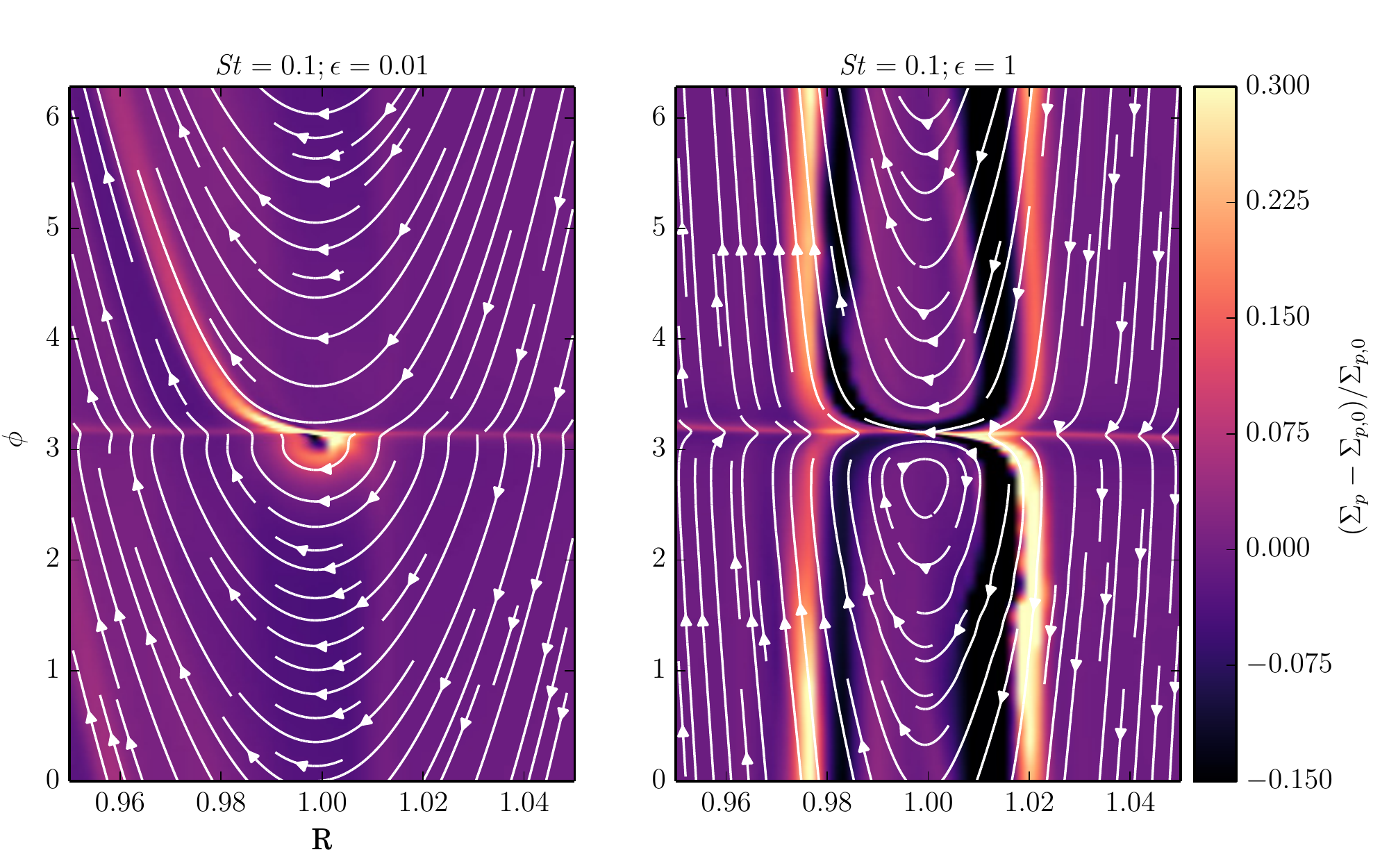}
\includegraphics[width=\columnwidth]{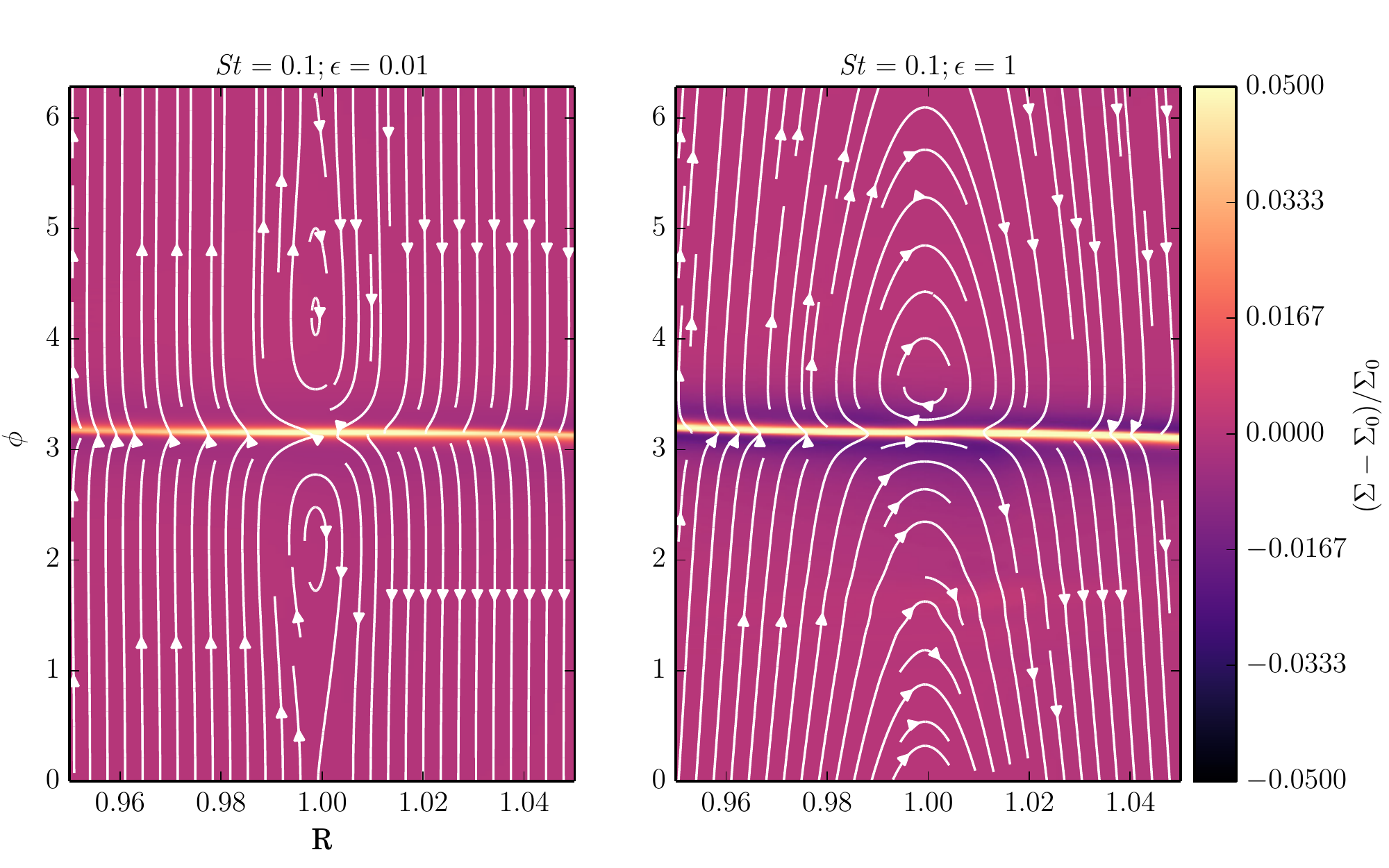}
\caption{{\it Top:} Streamlines and relative surface density perturbation of pebbles at Time=50 for $\st=0.1$ and for  solid-to-gas ratios of $\epsilon=0.01$ (Left) and $\epsilon=1$ (Right). {\it Bottom:}  Corresponding relative gas surface density perturbation and gas streamlines.}
\label{fig:backreaction}
\end{figure}

\begin{figure}
\centering
\includegraphics[width=\columnwidth]{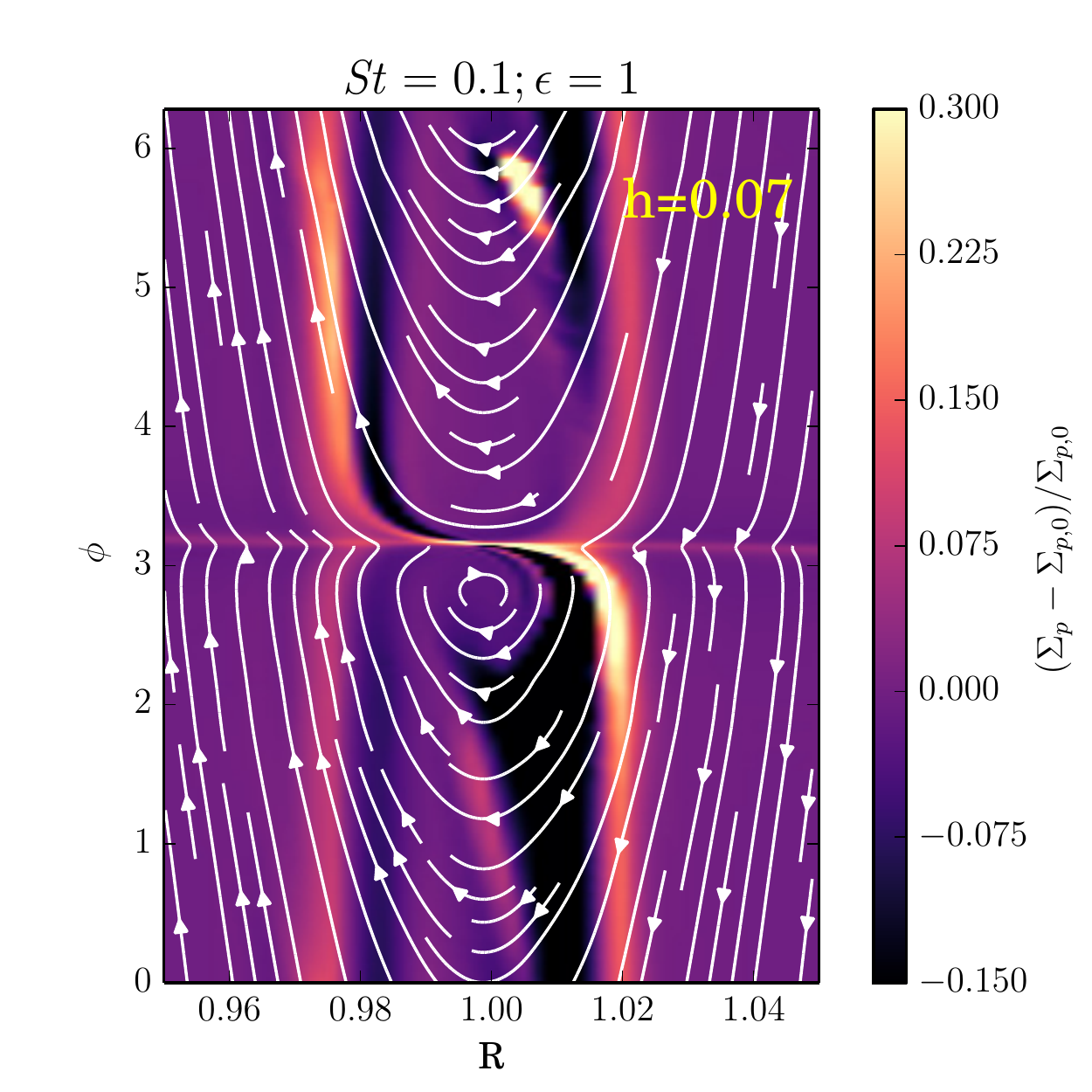}
\caption{Streamlines and relative surface density perturbation of pebbles at Time=50 for a simulation $\st=0.1$,  solid-to-gas ratio of $\epsilon=1$ corresponding to a thicker disc with $h\sim 0.07$. This model has therefore an effective scale height such that ${\tilde H}/R$=0.05.}
\label{fig:backreaction2}
\end{figure}

\begin{figure}
\centering
\includegraphics[width=0.49\columnwidth]{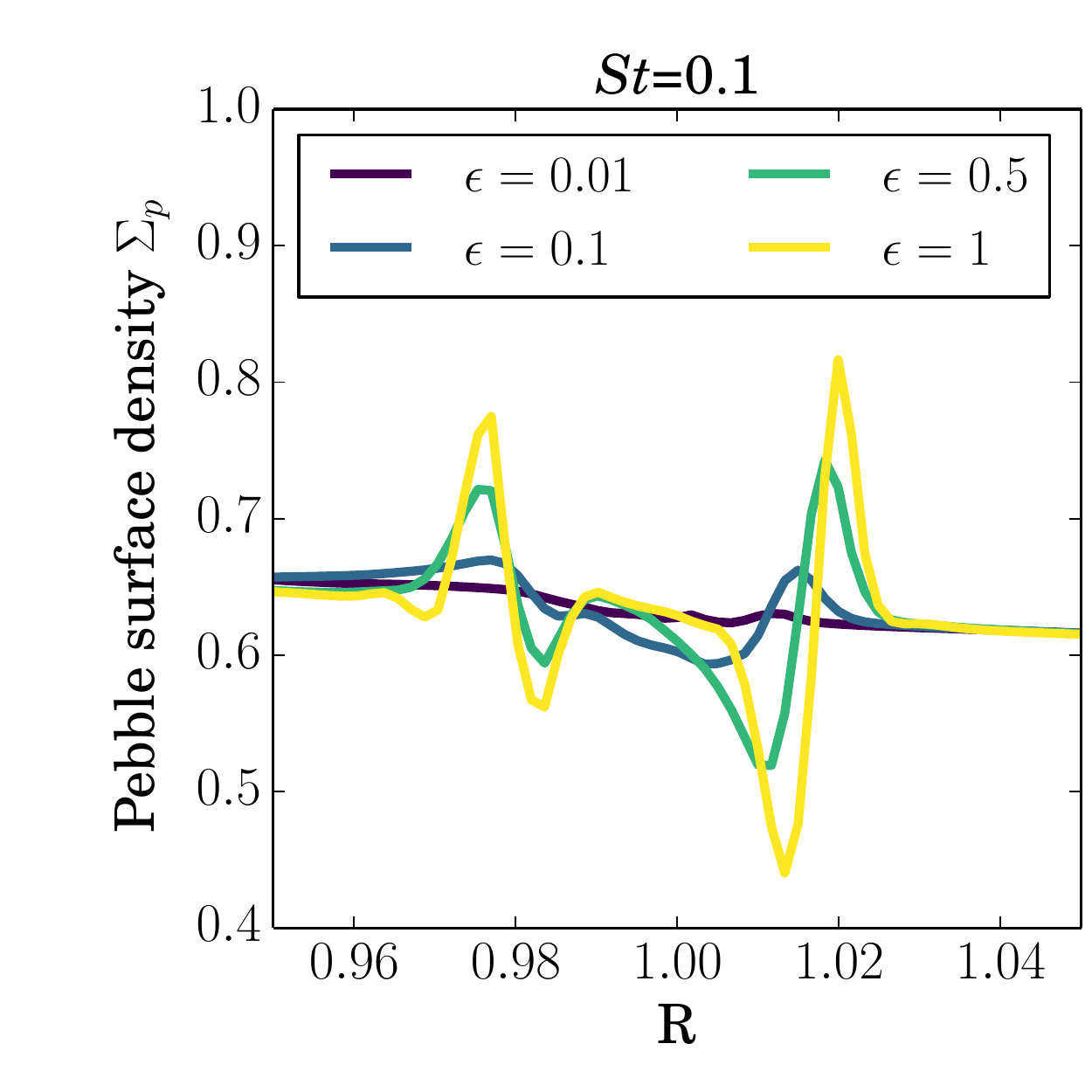}
\includegraphics[width=0.49\columnwidth]{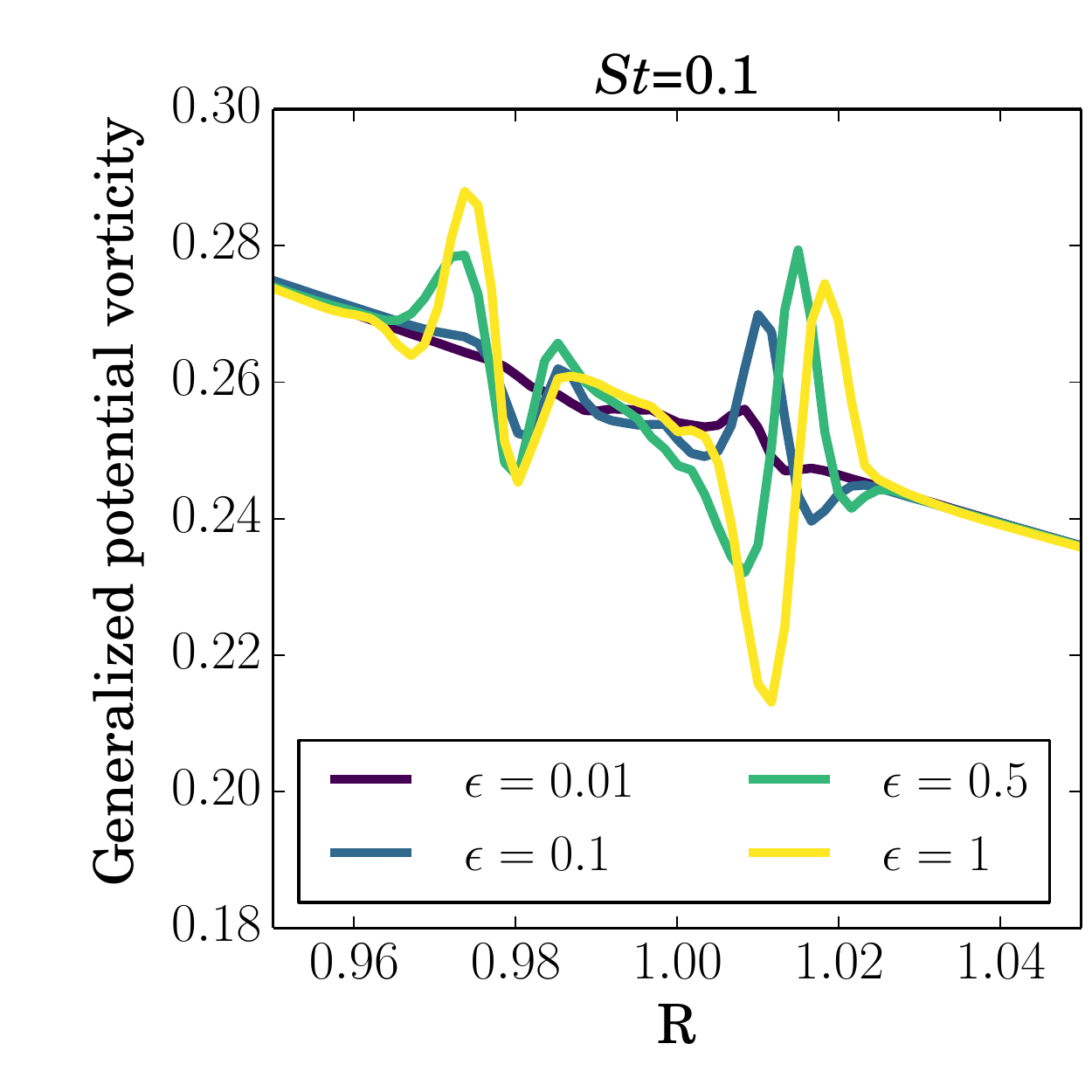}
\caption{{\it Left:}  Azimuthally averaged profile of the pebble surface density at Time=50 for $\st=0.1$ and for different values of the solid-to-gas  ratio $\epsilon$. {\it Right:} Corresponding  azimuthally averaged profiles of the generalized potential vorticity. }
\label{fig:gpv}
\end{figure}

We begin the discussion of the simulation results by describing the  impact of taking into account the effect of the particle backreaction onto the dynamics. For $\st=0.1$ particles,  we compare in Fig. \ref{fig:backreaction} the perturbed solid  and gas surface densities for the two cases $\epsilon=0.01$ and $\epsilon=1$. The  distribution of pebbles (upper panel) in the case with $\epsilon=0.01$ agrees fairly well with the results of  Ben\`itez-Llamblay \& Pessah (2018), revealing an overdense inner wake resulting from scattering in the vicinity of the planet;  together with an  overdense region just behind the planet, and inside which solids can accumulate. Pebble dynamics, however, is very different in the case with $\epsilon=1$ where the effects of the pebble backreaction are important. Here, overdense rings develop at the separatrix between some closed librating and circulating streamlines, as clearly revealed by the few streamlines that are overplotted in Fig. \ref{fig:backreaction}.   We remark that the pitch angle betwen the iso-contours of the pebble surface density and the streamlines can be quite large, which suggest that the streamlines can differ from the actual particle trajectories. Nevertheless,  close inspection of the upper right panel of Fig. \ref{fig:backreaction}  reveals that there is clear trend for the surface density at the downstream separatrix  to be higher than at the upstream separatrix. This results from the reduced drift rate, which makes most of the particles  pass very close to the planet before being scattered at the downstream separatrices. Such scattering of pebbles at the downstream separatrix can be also clearly identified in the hydrodynamical simulations including particles of Morbidelli \& Nesvorny (2012, see their Figure 5). The overdense stream at the front (resp. rear) side of the downstream separatrix then approaches the planet from the  rear (resp. front), created thereby the two overdense rings that are observed.   We remark in passing that for tightly coupled particles,  the sound speed is reduced due to dust loading which, by analogy with the classical gap-opening process in a gas disc,  may favor ring formation due to stronger non-linear effects.  To clearly isolate the effect of taking into account the effect of disc feedback, we illustrate in  Fig. \ref{fig:backreaction2} pebble dynamics resulting from a calculation with $h=0.07$ but where the smoothing length of the planet was kept similar to the case with $h=0.05$. For $\epsilon=1$, this model has effective aspect ratio ${\tilde h}=h/\sqrt{2}=0.05$, almost equivalent to the model with $h=0.05$ and $\epsilon=0.01$ and presented in top left panel of Fig. \ref{fig:backreaction}.  Ring formation is again clearly at work for this model, despite exhibiting weaker overdensities compared to the case with $h=0.05$ and $\epsilon=1$. This highlights the major role played by  the dust feedback and finite dust/gas coupling in ring formation.   \\

Turning back to Fig. \ref{fig:backreaction}, we see that, compared to the solid component, the gas distribution (lower right panel) remains relatively unperturbed. Another difference is that for the pebbles, the librating region is located at the rear of planet whereas for the gas, it is located in front of the planet. This simply occurs  because the gas flows outward as a result of the particle feedback.  \\

In a baroclinic gas disc, it is interesting to note that large entropy gradients can exist at the downstream  separatrices of a low-mass planet due to advection of entropy, giving rise to a corotation torque exerted on the planet (Masset \& Casoli 2009). In the inviscid limit, such large entropy gradients can also be responsible for the formation of vortices at these outgoing separatrices, and this has been reported in previous numerical simulations (Baruteau \& Masset 2008; Paardekooper et al. 2010). Vortex formation in that case results from the RWI,  which is expected to develop at entropy extrema of a gas disc. In a  dusty gas disc, we expect the RWI to be rather triggered at extrema of the dust-to-gas ratio $\epsilon$ because the effective entropy of a gas$+$dust mixture is a function of  $\epsilon$, as shown by Lin \& Youdin (2017). From the above discussion, we therefore expect the separatrix to be a favored location for the RWI as a result of ring formation. More precisely, it is expected the development of the RWI to occur  at extrema of the generalized potential vorticity (PV),  which is defined as (Lin \& Youdin 2017):

\begin{equation}
{\mathcal V}=\frac{\kappa^2}{2\Omega (\Sigma+\Sigma_p)}\left(1+\frac{\Sigma_p}{\Sigma}\right)^2
\end{equation}

 Fig. \ref{fig:gpv} shows  the profiles of the pebble surface density and generalized PV at Time=50
for various values of the initial solid-to-gas ratio. We see that there is a clear trend for the amplitude of the pebble density bump to increase with solid-to-gas ratio, as a consequence of a  weaker gas drag.  A shallow  dip in the pebble surface density profile is also clearly visible, and which results from the loss through radial drift of  material located in between the two rings.  The locations of the two pebble rings correspond to the two  maxima that are present in the generalized PV profile, and which therefore appear as favored locations for the growth of the RWI. 

\subsection{Evolution for $\st=0.1$ particles}
\label{sec:st0v1}

\begin{figure*}
\centering
\includegraphics[width=\textwidth]{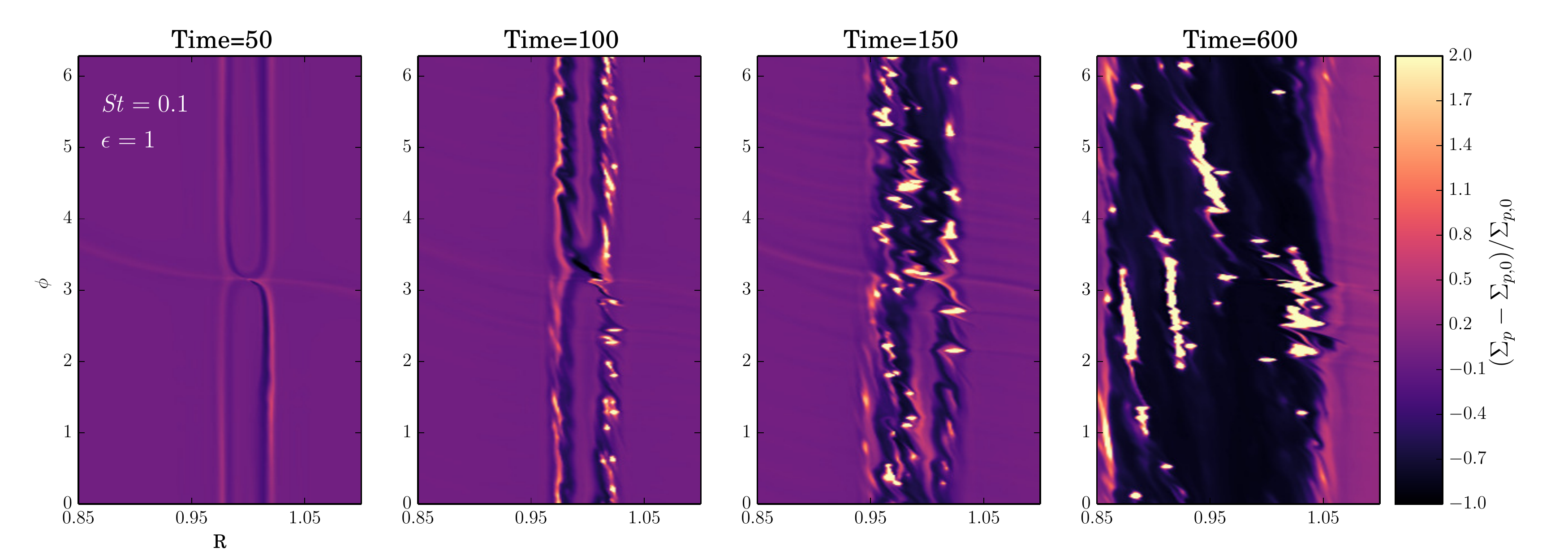}
\includegraphics[width=\textwidth]{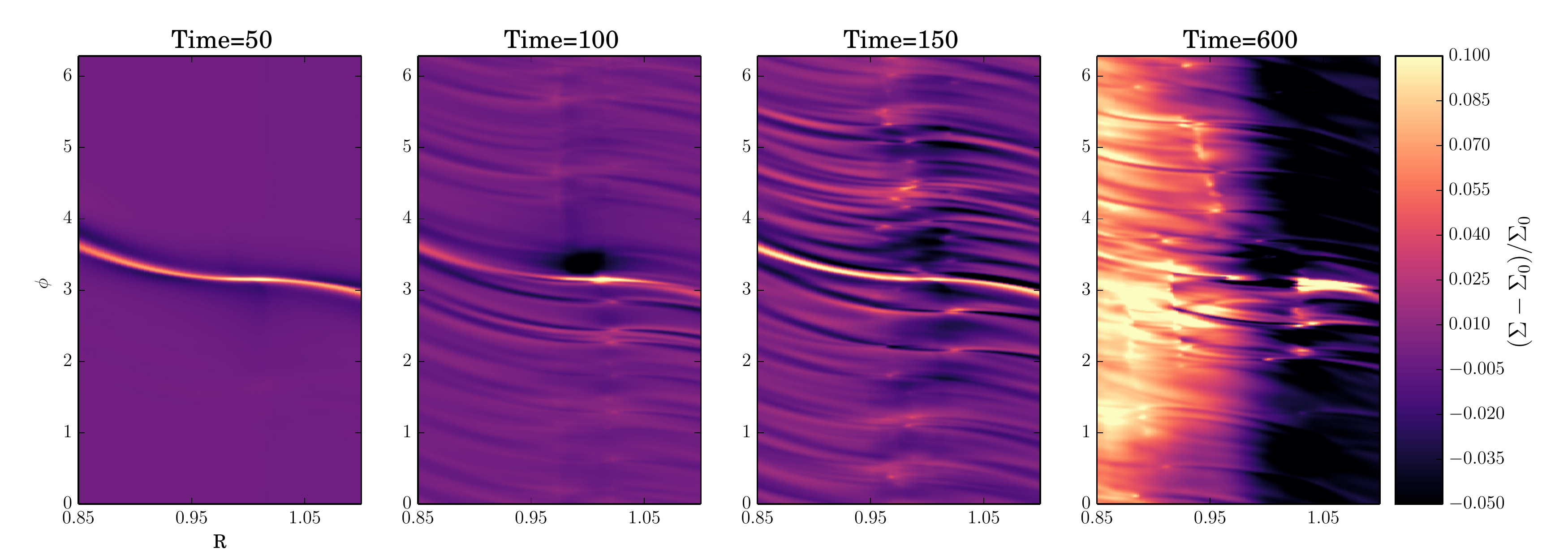}
\includegraphics[width=\textwidth]{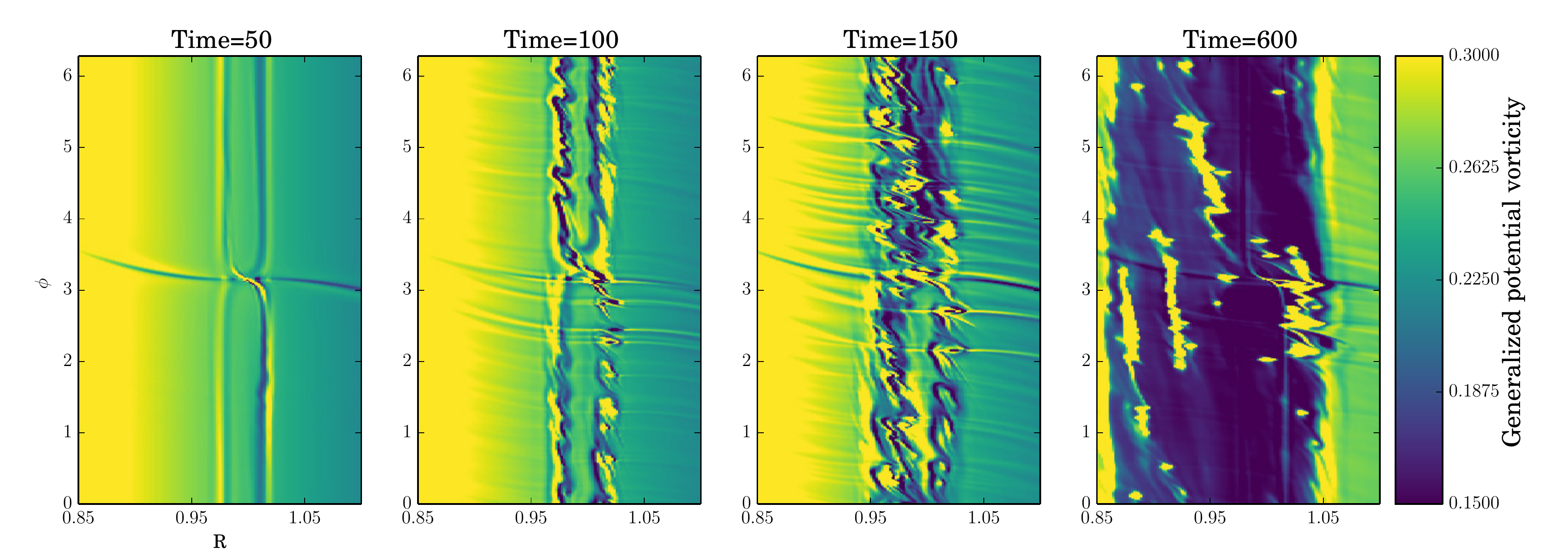}
\caption{{\it Top row}: Evolution of the relative pebble surface density perturbation for the run with ${\st=0.1}$ and $\epsilon=1$. {\it Middle row:}  Evolution of the relative gas surface density pertubation. {\it Bottom row:} Evolution of the generalized potential vorticity. }
\label{fig:fiducial}
\end{figure*}

\begin{figure}
\centering
\includegraphics[width=\columnwidth]{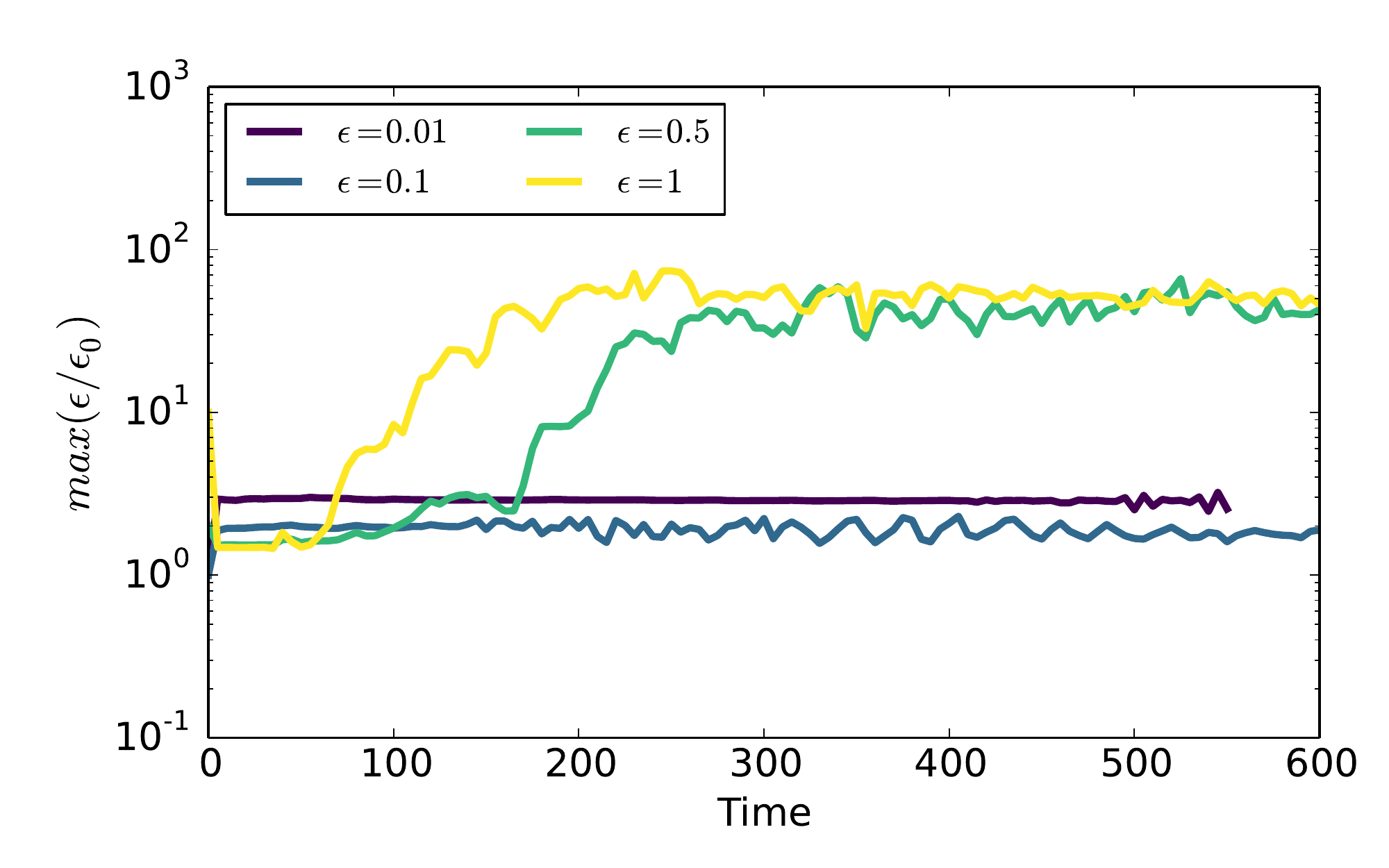}
\includegraphics[width=\columnwidth]{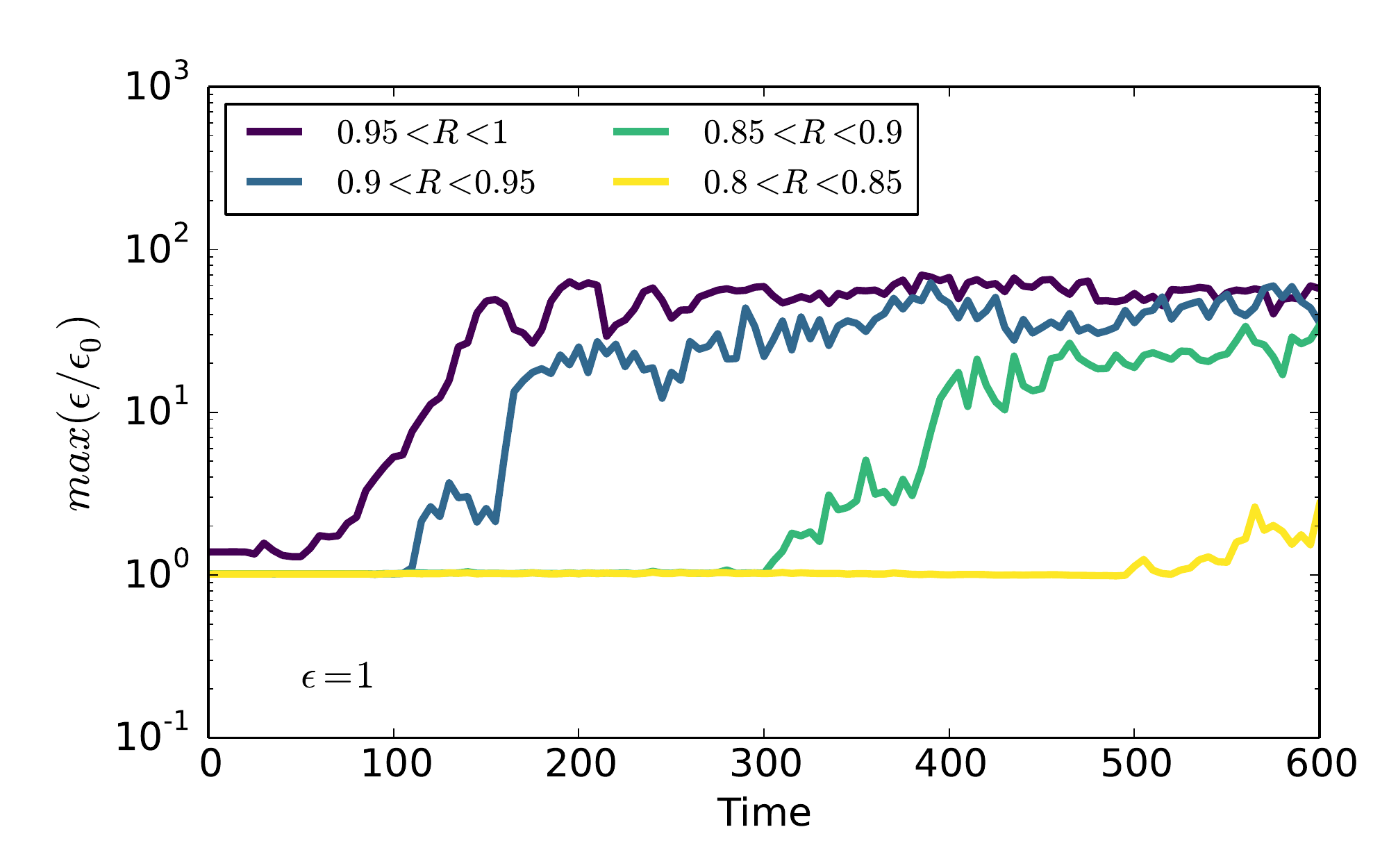}
\caption{{\it Top}: Maximum solid-to-gas ratio $\epsilon_{max}$ as a function of time, relative to the initial solid-to-gas-ratio $\epsilon_{0}$, for $\st=0.1$ and for different values for $\epsilon$.  {\it Bottom:} Time evolution of $\epsilon_{max}$ for $\st=0.1$ and $\epsilon=1$ at different radius bins. It shows that the instability propagates inward because of the combined effect of particle trapping and drift.}
\label{fig:ratiomax}
\end{figure}

For the fiducial simulation with $\st=0.1$ and $\epsilon=1$, Fig. \ref{fig:fiducial} shows the evolution of the  dust (top panel) and gas (bottom panel) components.  Using Eq. \ref{eq:vrd}, the radial mass flux of solids at the location of the planet is approximately given by ${\dot M_p}=2\pi RV_R\Sigma_p\sim\pi R \Sigma_p \st  \eta v_k$ for this model, which would correspond to ${\dot M}\sim 0.03\;M_\oplus/$yr  in our units. Consistently with the previous discussion about the RWI in dusty discs, this instability is found to develop at the planet separatrix, with multiple dusty eddies forming at early times.  We remark that in the pure gas case without self-gravity, the most unstable azimuthal wavenumber for the RWI is $m\sim3-5$ whereas much higher-m is favored in the dusty case here. Also, the length scales  appear  much smaller than the gas scale height, which is probably due to the sharp gradients that are present in the generalized PV profile (see right panel in Fig. \ref{fig:gpv}). Interestingly,  the gas remains only  weakly impacted during the early growth stage of the instability, although small-scale vortices can be distinguished in the panel corresponding to the gas distribution at Time=100.\\

 For this run  and for different values of the metallicity as well, we show in the upper panel of Fig. \ref{fig:ratiomax}  the time evolution of the maximum of the solid-to-gas ratio $\epsilon_{max}$. For $\epsilon\ge 0.5$, this figure indicates  exponential growth for  $\epsilon_{max}$, before it saturates to  $\epsilon_{max}\sim 50$. Increase in dust-to-gas ratio can occur because small-scale vortices  correspond to local pressure maxima that might therefore be able to further trap pebbles.  A process  that can also lead to an increase in $\epsilon_{max}$ is related to the fact that  dusty eddies with lower solid-to-gas ratio drift more rapidly in the azimuthal direction and can catch up with vortices with slightly higher solid-to-gas ratio. In that sense, the mecanism that is at work here can be similar to what happens in the context of the streaming instability, where dust clumps are able to collect more inward-drifting dust particles.  This is suggested by comparing the Time=100 and  Time=150 panels in the top row of Fig. \ref{fig:fiducial}, where we see that dusty clumps tend to concentrate into larger-scale structures.  The reorganization of dusty clumps into large structures is also evident when examining the Time=600 panel. At that time,  dusty eddies have concentrated into dense filaments,  which might eventually form bound clusters of particles in presence of self-gravity. Again, the formation of such azimuthally extended structures is a generic outcome of the streaming instability (Johansen \& Youdin 2007) or the Kelvin-Helmoltz instability with $\st=1$ dust particles (Johansen et al. 2006).\\

The lower panel of  Fig. \ref{fig:ratiomax} shows for the fiducial run the evolution of $\epsilon_{max}$  at different radius bins. It reveals that the instability at the separatrix tends to propagate inward. This arises due the combined effect of clumping plus radial drift which causes  the inner edge of the gap to progressively migrate inward. Turning back to the Time=600 panel in Fig. \ref{fig:fiducial}, this feature can  be clearly observed by inspecting the  distribution of solids at that time. Here, the inner edge of the pebble gap appears to be located at $R\sim 0.85$ whereas the outer edge of the gap is at $R\sim 1.05$.  Such an asymmetric gap would have potential important consequences on the dust torques felt by the planet. In can be indeed reasonably expected that the outer disc would provide the main contribution to the torques in that case, resulting in the inward migration of the planet. There is also a slight trend for the outer edge of the gap to move outward, although much more slowly because of particles drifting inward from the outer disc. \\
Regarding the gas component, an interesting feature is that at late times, the gas surface density is much higher inside the orbit the planet, whereas  the outer disc  tends to be gas-depleted (see middle-right panel of Fig.  \ref{fig:fiducial}) . This occurs because within the pebble gap, the feedback onto the gas is reduced so that the outward drift velocity of the gas there  is smaller than the velocity of the gas located outside of the gap. In the inner (resp. outer) disc, this consequently results in an accumulation (resp. depletion) of gas at the inner (resp. outer) edge of the gap.

\subsection{Evolution as a function of Stokes number}
\label{sec:stokes}

\begin{figure}
\centering
\includegraphics[width=\columnwidth]{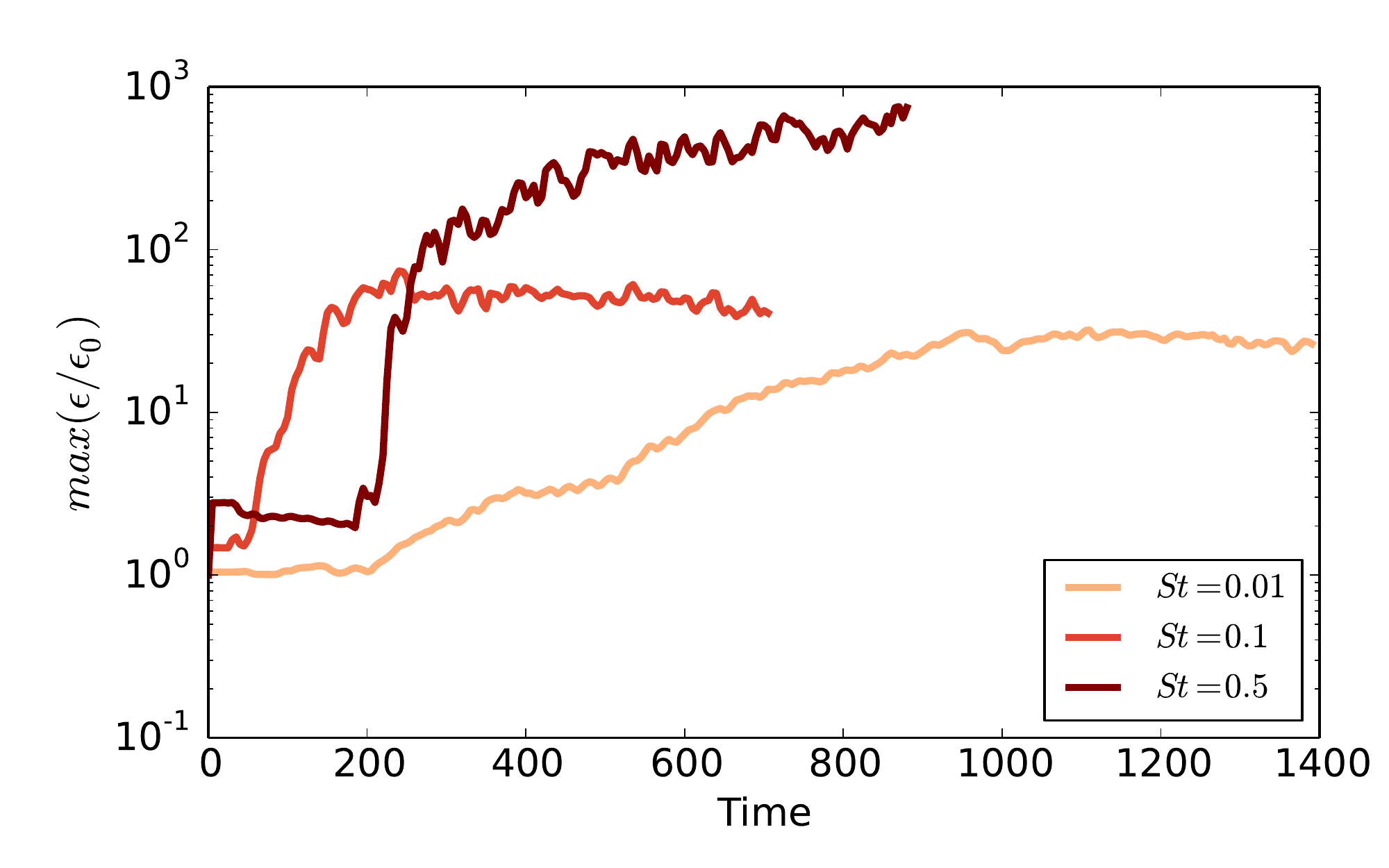}
\caption{Maximum solid-to-gas ratio $\epsilon_{max}$ as a function of time, relative to the initial solid-to-gas-ratio $\epsilon_{0}$, for $\epsilon=1$ initially and for different values of the Stokes number $\st$. }
\label{fig:ratiostokes}
\end{figure}

\begin{figure*}
\centering
\includegraphics[width=\textwidth]{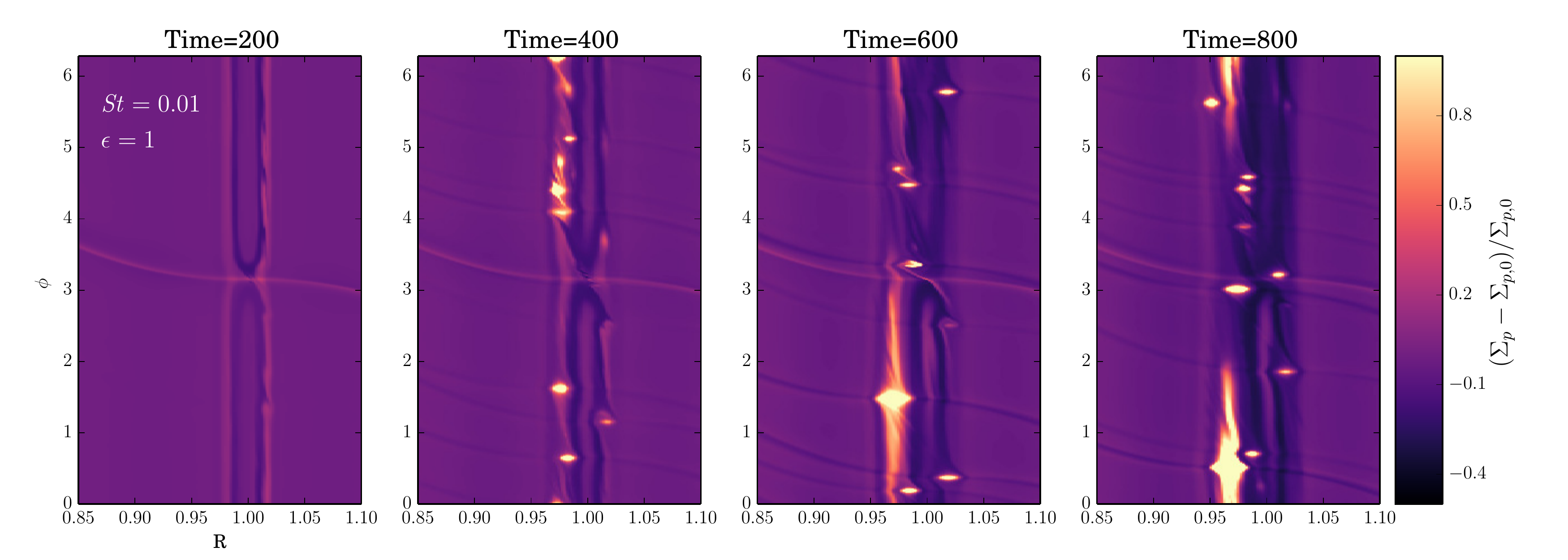}
\includegraphics[width=\textwidth]{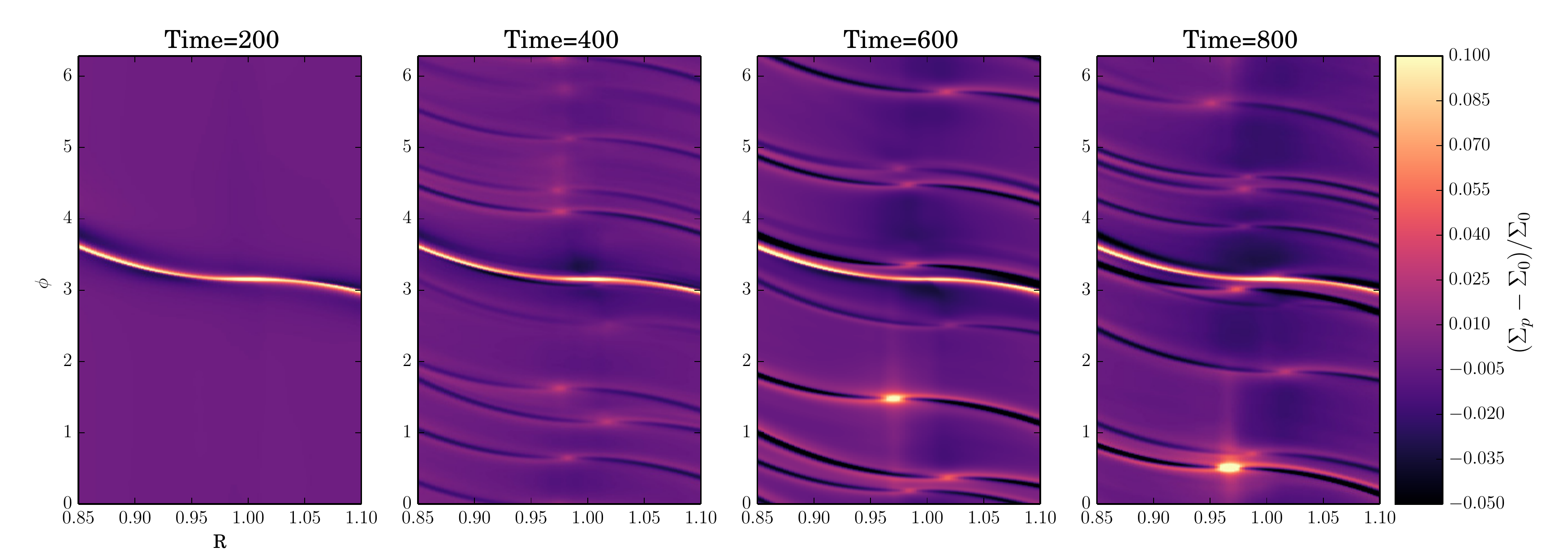}
\caption{{\it Top row}: Evolution of the relative pebble surface density perturbation for the run with ${\st=0.01}$ and $\epsilon=1$. {\it Bottom row:}  Evolution of the relative gas surface density pertubation.}
\label{fig:st0v01}
\end{figure*}

\begin{figure}
\centering
\includegraphics[width=\columnwidth]{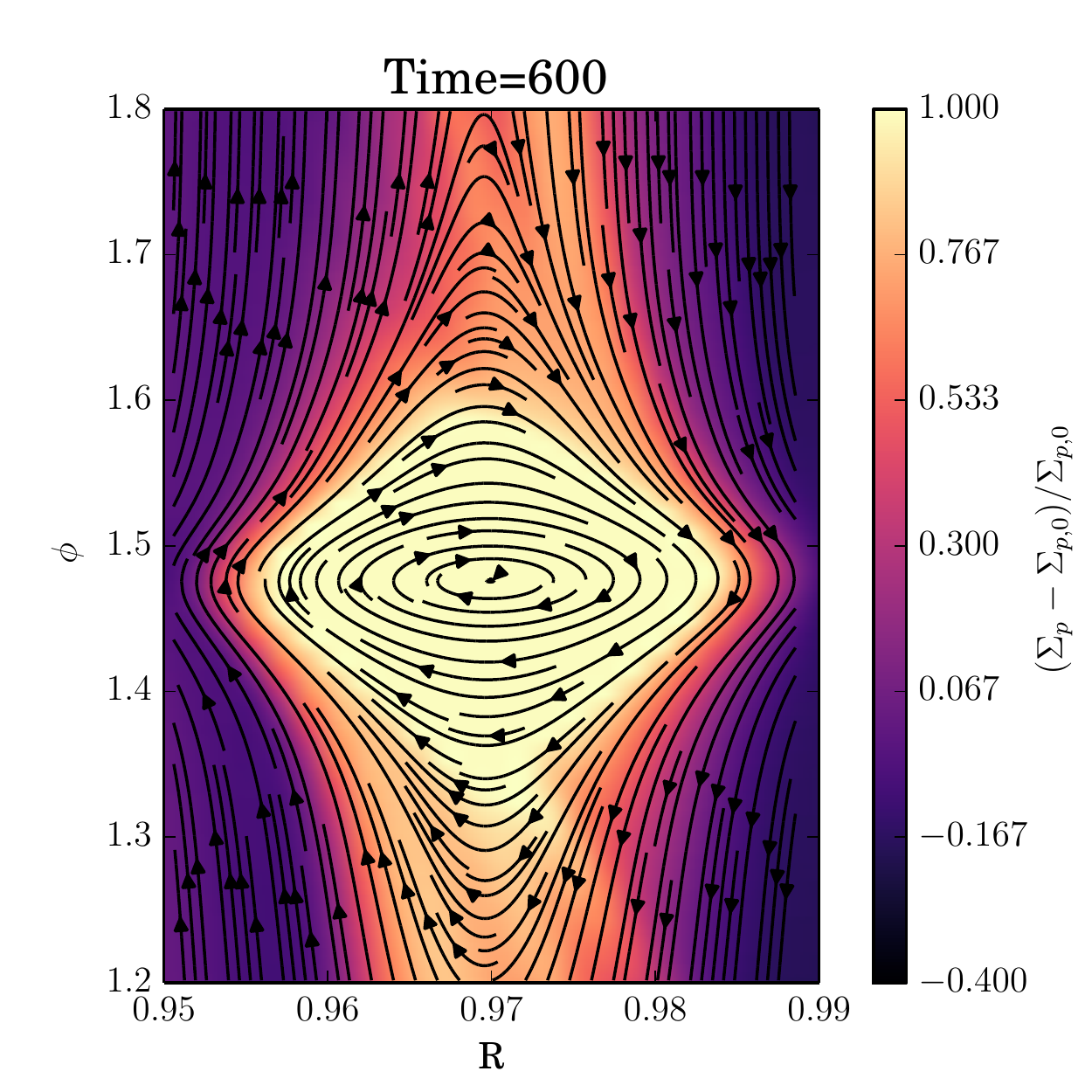}
\caption{Zoom on the vortex that is formed at Time=600 for the run with ${\st=0.01}$ and $\epsilon=1$ (see third panel in Fig. \ref{fig:st0v01}) together with streamlines of pebbles (black lines) within the vortex.}
\label{fig:streamlines_st0v01}
\end{figure}

\begin{figure}
\centering
\includegraphics[width=0.49\columnwidth]{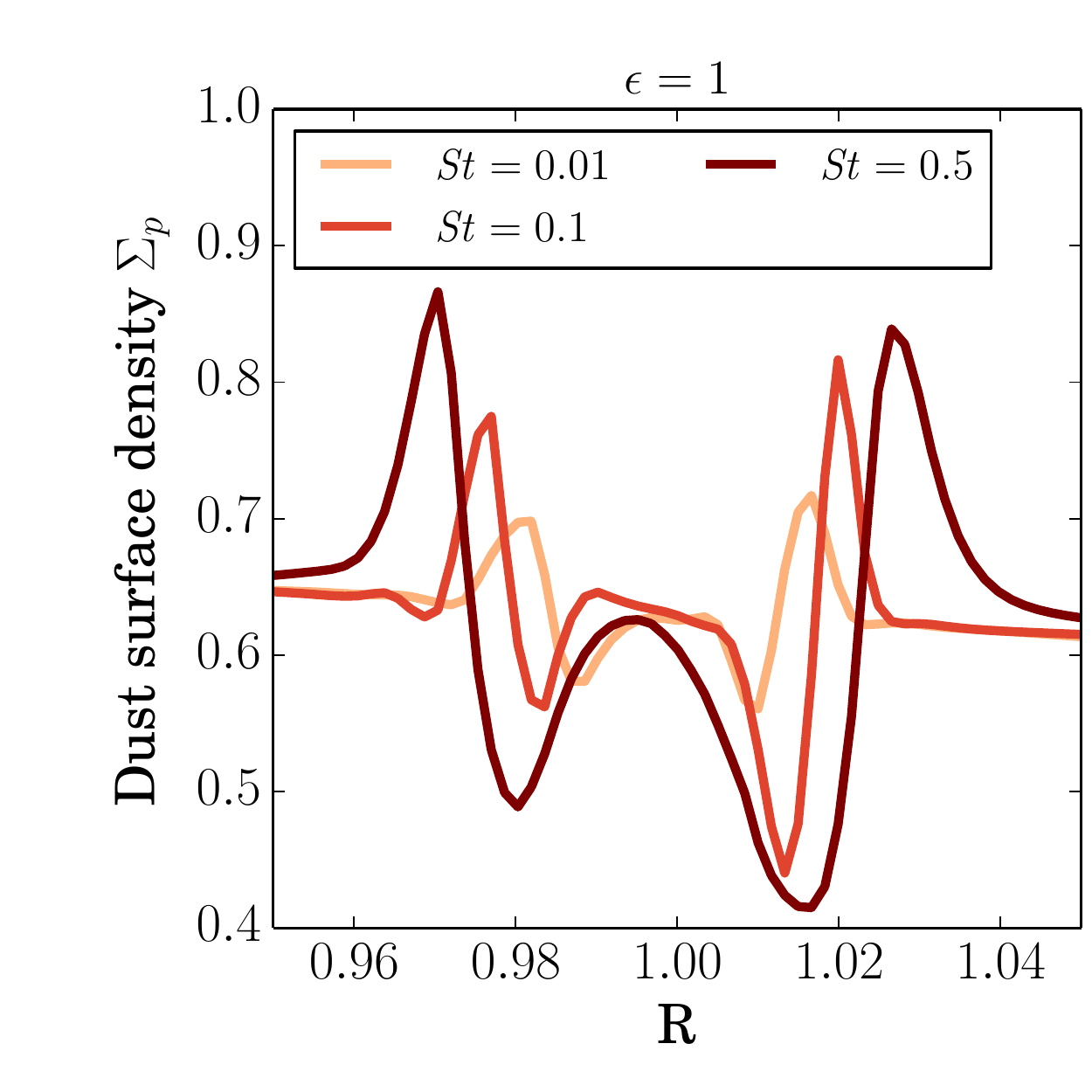}
\includegraphics[width=0.49\columnwidth]{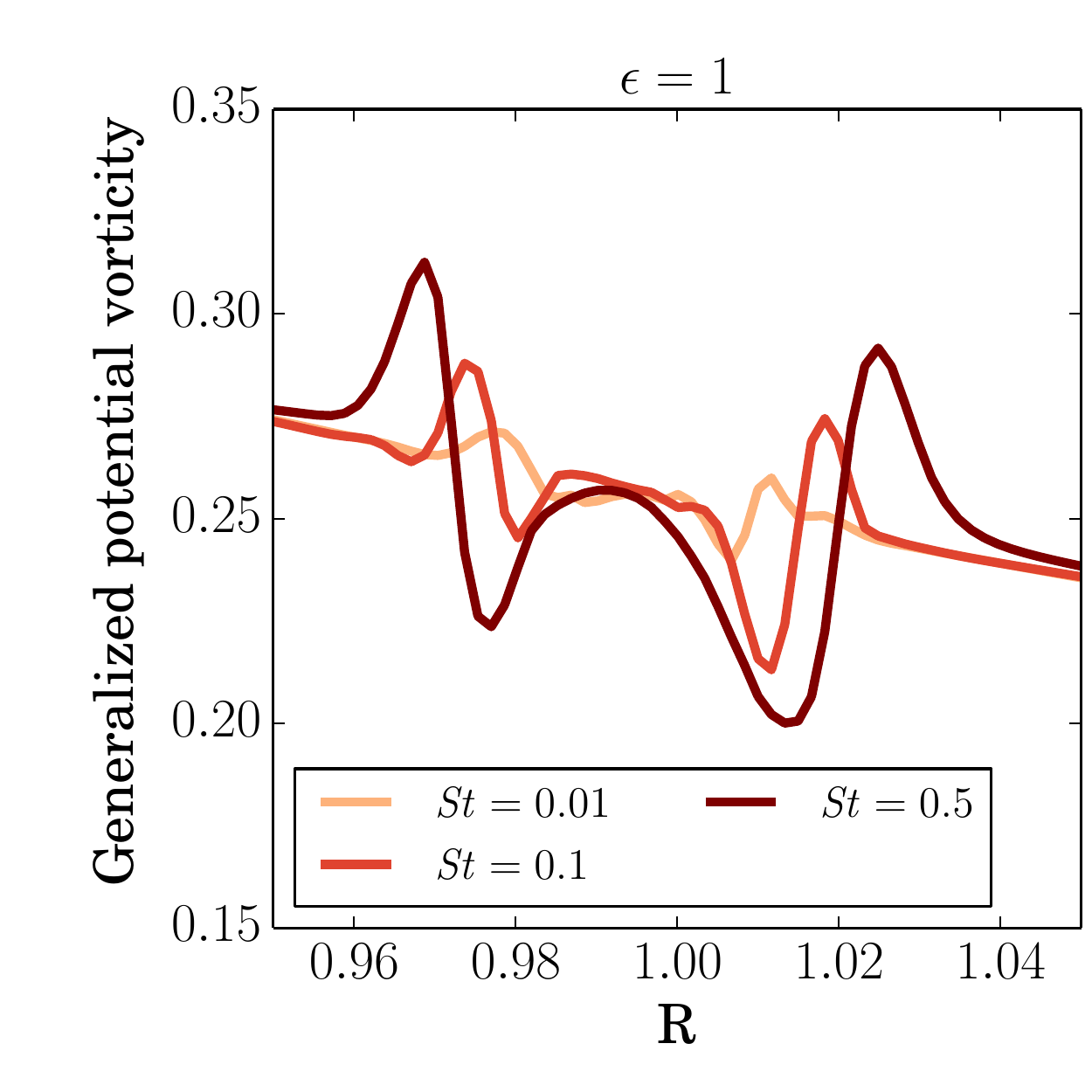}
\caption{{\it Left:}  Azimuthally averaged profile of the pebble surface density at Time=50 for $\epsilon=1$ initially and for different values of the Stokes number $\st$. {\it Right:} Corresponding  azimuthally averaged profiles of the generalized potential vorticity.  For clarity, each quantity has been normalized by its value at the inner edge of the disc. }
\label{fig:gpvstokes}
\end{figure}

\begin{figure}
\centering
\includegraphics[width=0.49\columnwidth]{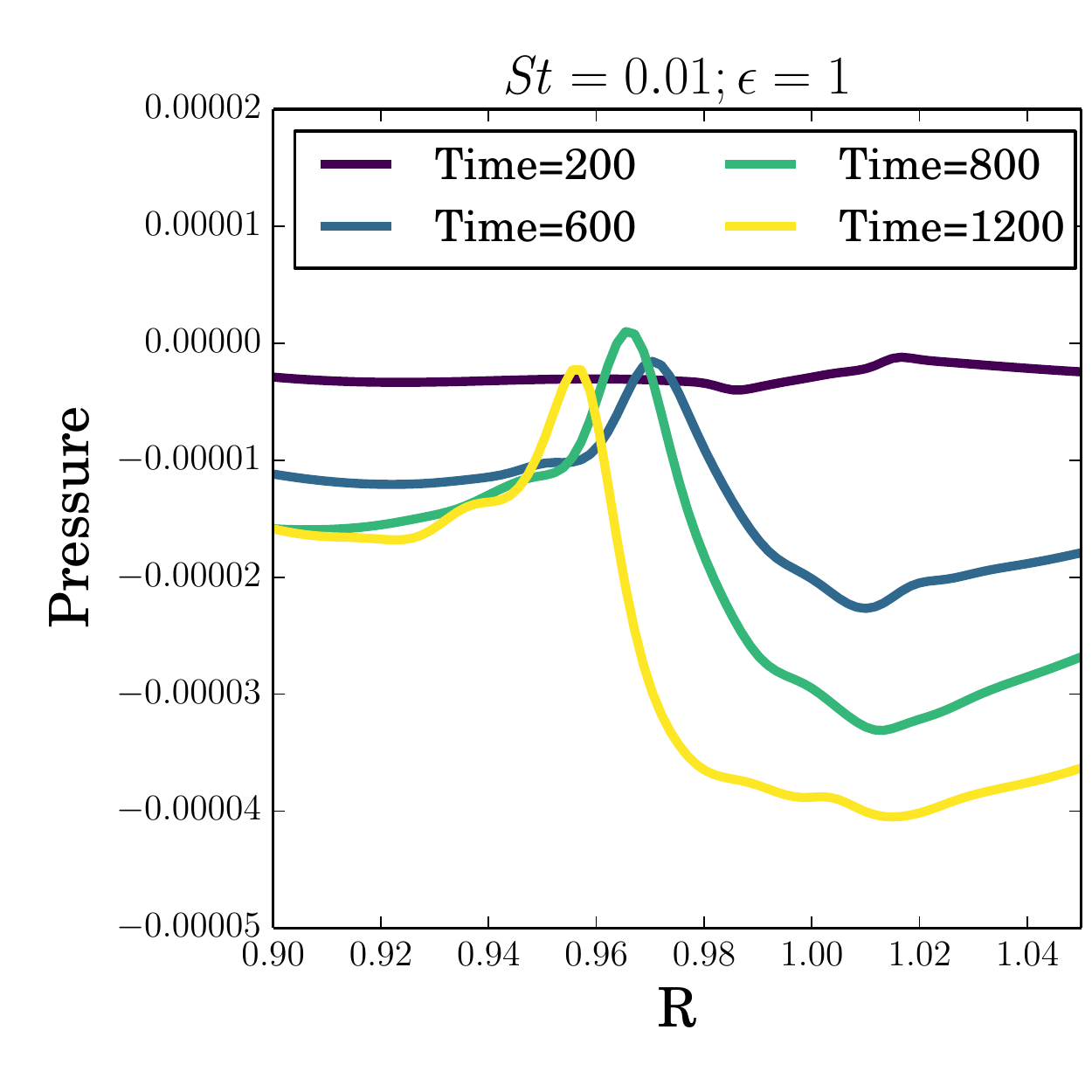}
\includegraphics[width=0.49\columnwidth]{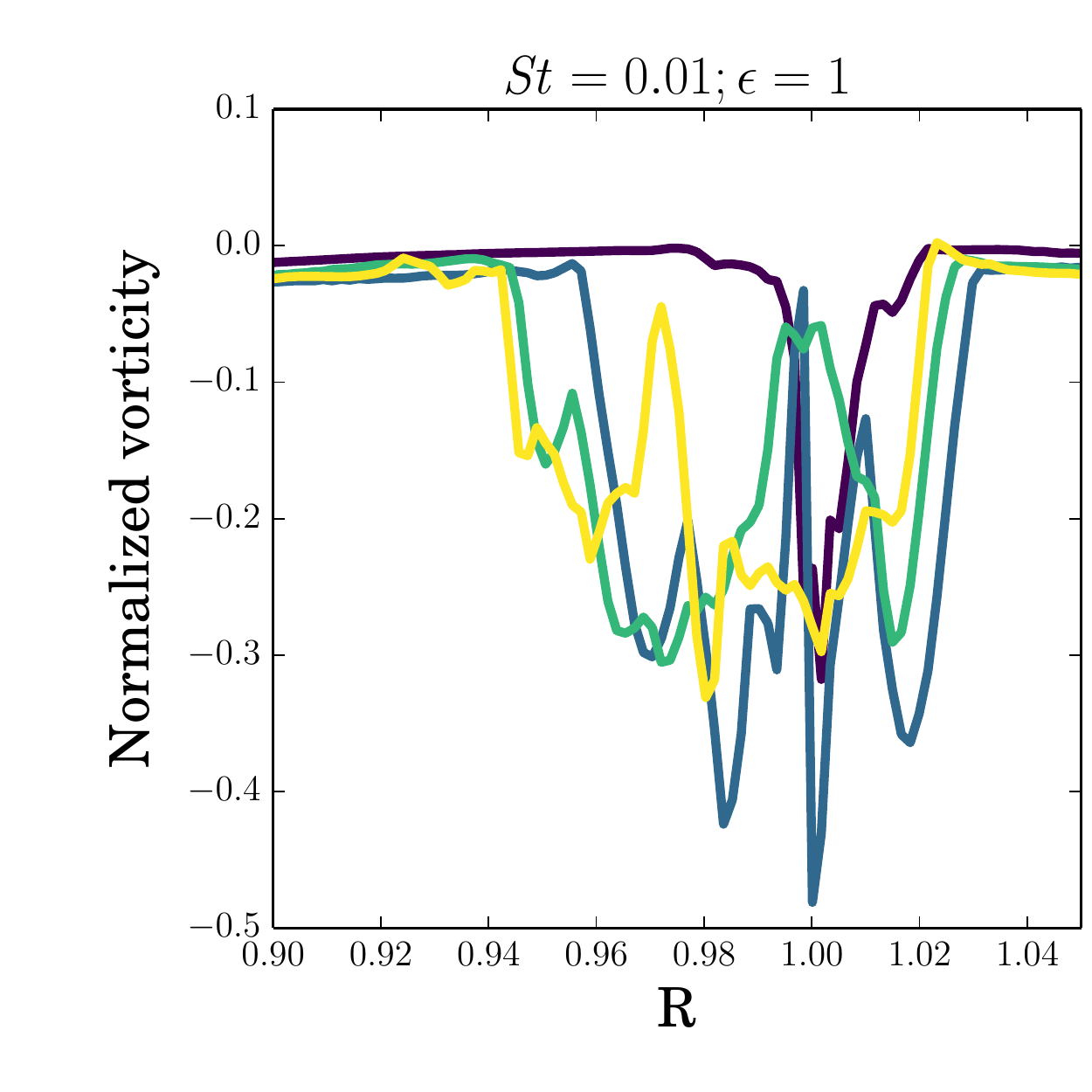}
\includegraphics[width=0.49\columnwidth]{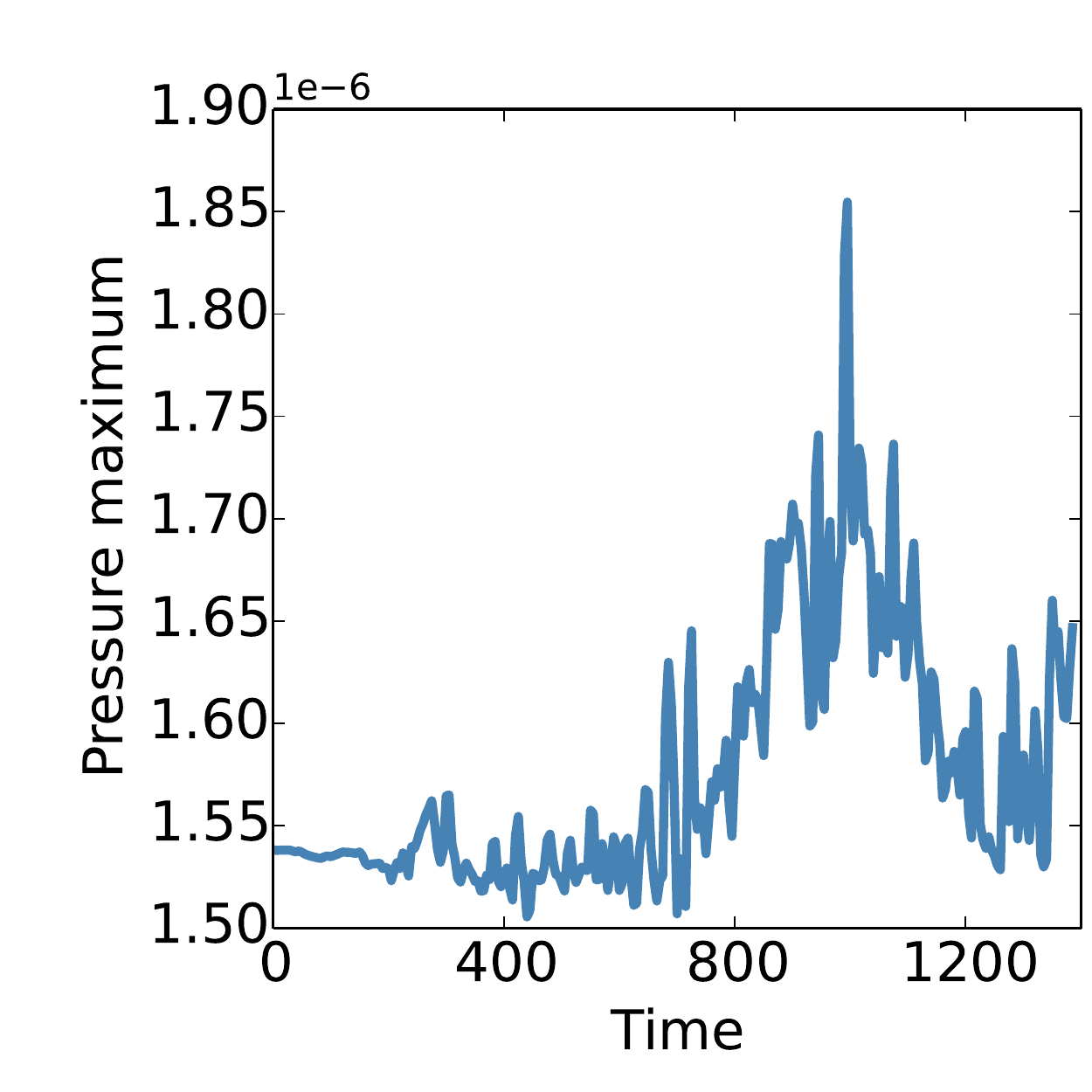}
\includegraphics[width=0.49\columnwidth]{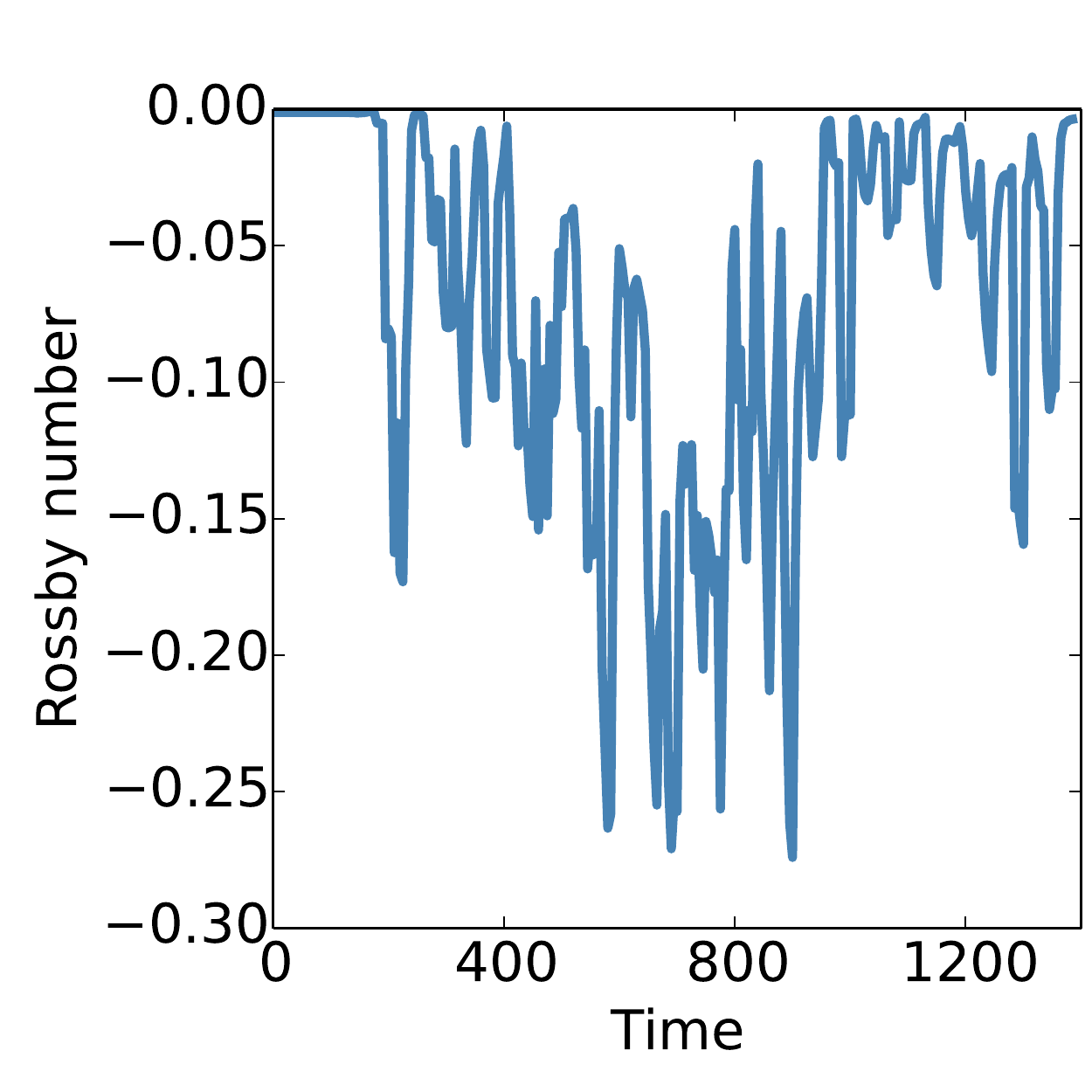}
\caption{{\it Top:}  Azimuthally averaged profile of the gas pressure at different times for $\epsilon=1$ initially and  Stokes number $\st=0.01$ ({\it Left}) and azimuthally averaged profile  of the normalized gas vorticity at the sames times ({\it Right}). {\it Bottom:}  For the same run, time evolutions of the pressure maximum at vortex center  ({\it Left}) and Rossby number   ({\it Right}).  }
\label{fig:saturate}
\end{figure}

\begin{figure}
\centering
\includegraphics[width=0.49\columnwidth]{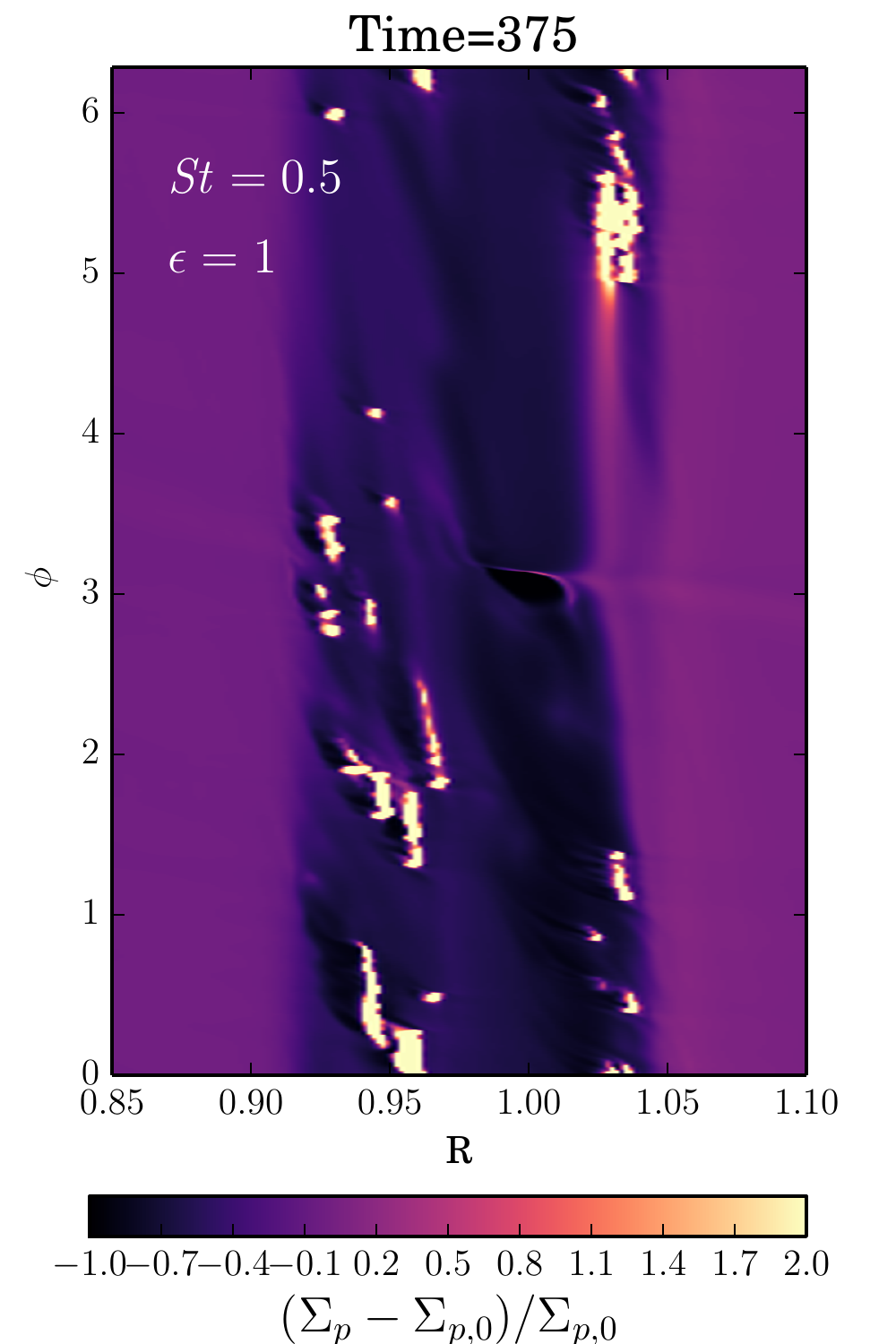}
\includegraphics[width=0.49\columnwidth]{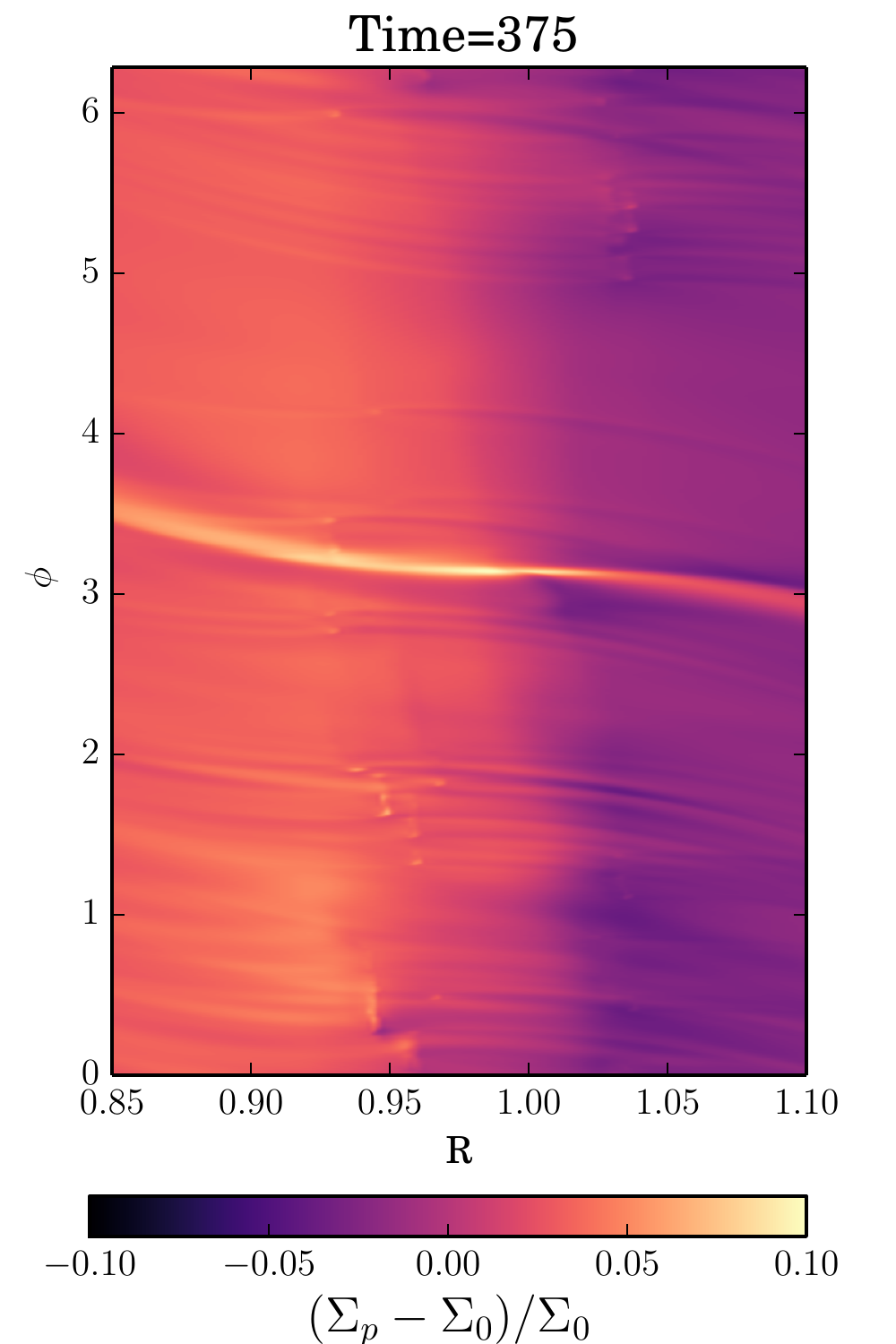}
\caption{{\it Left}: Relative pebble surface density perturbation at Time=375 for the run with ${\st=0.5}$ and $\epsilon=1$. {\it Right:}  Corresponding relative gas surface density pertubation.}
\label{fig:st0v5}
\end{figure}

In this section we present the results of simulations with $\st=0.01, 0.5$ aimed at examining  how the linear and saturated stages of the instability depend on the Stokes number. The maximum value of the solid-to-gas ratio $\epsilon_{max}$ as a function of time is shown in Fig. \ref{fig:ratiostokes}. For each model, exponential growth of $\epsilon_{max}$ is observed, with an associated growth rate that increases with Stokes number. This is not surprising since it is expected pebble  capture to be more efficient  for particles with $\st\sim1$ that undergo significant gas drag than for particles with $\st\sim 0.01$ that are more coupled to the gas. For $\st=0.01$, the saturated value for $\epsilon_{max}$ is very similar to that found in the run with $\st=0.1$,  although a different mode of evolution was found in that case, as can be observed in the sequence shown in Fig. \ref{fig:st0v01}. Compared to the simulation with $\st=0.1$, the Time=400 panel of the pebble surface density reveals  that only a few vortices are formed at the planet separatrix, but with a larger radial size. Extrapolating down to smaller values of the Stokes number, we would expect  that  a large, single vortex to be eventually formed in the limit of small dust grains  with $\st \rightarrow 0$, which would be consistent with the results of Chen \& Lin (2018). From the pebble surface density  profile that is plotted for each model in the left panel of Fig.\ref{fig:gpvstokes}, it appears that  this arises because the width of the separatrix tends to increase as the Stokes number decreases, resulting in a weaker bump in the generalized PV profile (see right panel of Fig. \ref{fig:gpvstokes}).  We note that increase in the separatrix width is associated with a decrease in the  half-width of the horseshoe region $x_s$. For $\st=0.01$ , $x_s\sim \sqrt{q/h}\sim 0.01$ whereas we find  for $\st=0.5$ $x_s\sim 2.5R_H\sim 0.03$, where $R_H$ is the planet Hill Radius, which is consistent with the values reported by Benitez-Llambay \& Pessah (2018). As a result of a larger separatrix width, the evolution of the system for $\st=0.01$ is in some way less violent compared to the case with $\st=0.1$, involving in particular the formation of  a massive vortex that traps pebbles until $\epsilon_{max}\sim 50$. We see that dust capture at vortex center leads to an exponential growth in $\epsilon_{max}$, consistently with the results of  Surville et al. (2016).  The dynamics of pebbles in the vicinity of the vortex is illustrated in Fig. \ref{fig:streamlines_st0v01}. Crossing orbits seem not to play an important role for this value of the Stokes number, which validates our two-fluid approach, at least for this case. Here,  saturation  arises once the amount of trapped particles is high enough to make the vorticity inside the vortex decrease, causing subsequently the local capture of particles to become less efficient. This is illustrated in the top panel of Fig. \ref{fig:saturate} where are plotted the azimuthally averaged pressure and vorticity  profiles at different times.  We see that the pressure bump at the location of the vortex is not significantly altered despite the accumulation of particles, which implies that the outward gas flow due to particle feedback does not significantly flatten the pressure bump (Taki et al. 2016), at least at early evolution times. However, the upper right panel of Fig. \ref{fig:saturate} suggests that the main consequence of  dust trapping is a decrease in the vortex vorticity.  This can be confirmed by examining the joined time evolution of the pressure maximum and Rossby number $Ro$ at vortex center, with:
\begin{equation}
Ro=\frac{{\bf e_z}\cdot(\nabla\wedge({\bf v}-R\Omega_k{\bf e_\phi}))}{2\Omega_k}
\end{equation}
and which are displayed in the bottom panel of  Fig. \ref{fig:saturate}. The rossby number initially decreases as the vortex strengthens, until it reaches a  minimum value of $Ro\sim -0.2$  at Time=800, and  is then observed to weaken. However,  it is interesting to note that the pressure maximum still increases between Time=800 and Time=1000 before subsequently decreasing. The late increase in the pressure maximum that can be seen at Time=1200 suggests that this cycle can repeat at later times.\\

The simulation outcome for $\st=0.5$ is similar to the case $\st=0.1$,  with filaments of particles forming in the pebble gap while gas accumulates inside of the planet orbit. This is illustrated by the contours of  solid and gas surface densities at Time $=375$ in Fig. \ref{fig:st0v5}.  From the time evolution of $\epsilon_{max}$ in Fig. \ref{fig:ratiostokes},  we see that the main difference is that much higher values for $\epsilon_{max}$ are reached for $\st=0.5$, with a value of $\epsilon_{max}\sim 800$ obtained at the end of the run.  It is interesting to notice that similar values of $\epsilon_{max}$ have been reported in numerical simulations of the streaming instability (Johansen \& Youdin 2007; Bai  \& Stone 2010).
A higher value for $\epsilon_{max}$ in the case with $\st=0.5$ is not too much  surprising since i) the effect of gas drag is almost optimal for this value of the Stokes number and ii) the generalized PV peak at the separatrix and which is responsible for seeding the instability is more pronounced in the case with $\st=0.5$ (see Fig. \ref{fig:gpvstokes}).   We note that we also performed an additional simulation with $\st=5$, although the approximation of a pressureless fluid may break down for such a high value of the Stokes number (Hersant 2009). Neither onset of the instability nor vortex formation occured  in that case, which suggests that solid particles need to be moderately coupled to the gas for the instability to be triggered. 

 \subsection{Effect of resolution}

\begin{figure}
\centering
\includegraphics[width=0.49\columnwidth]{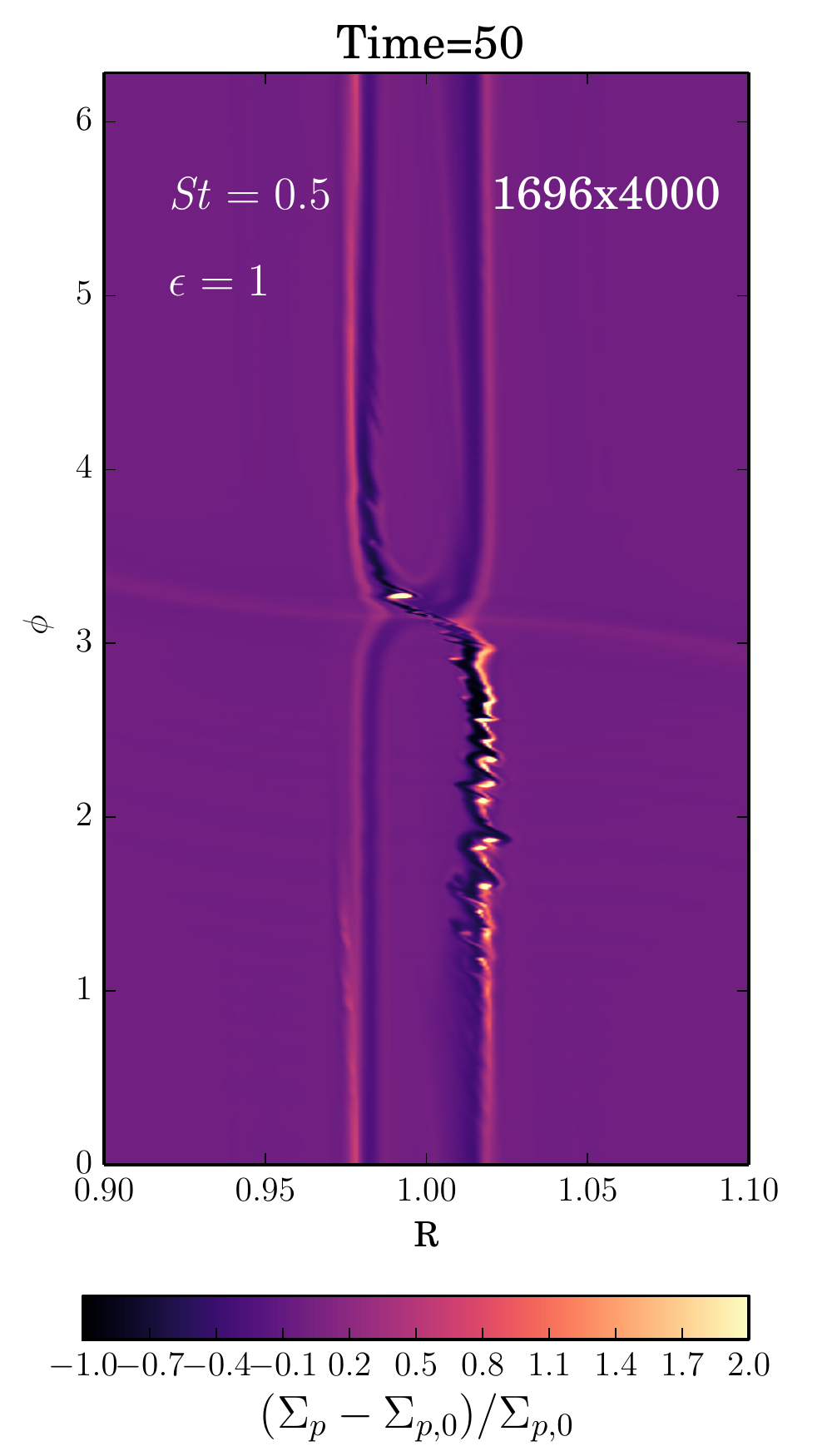}
\includegraphics[width=0.49\columnwidth]{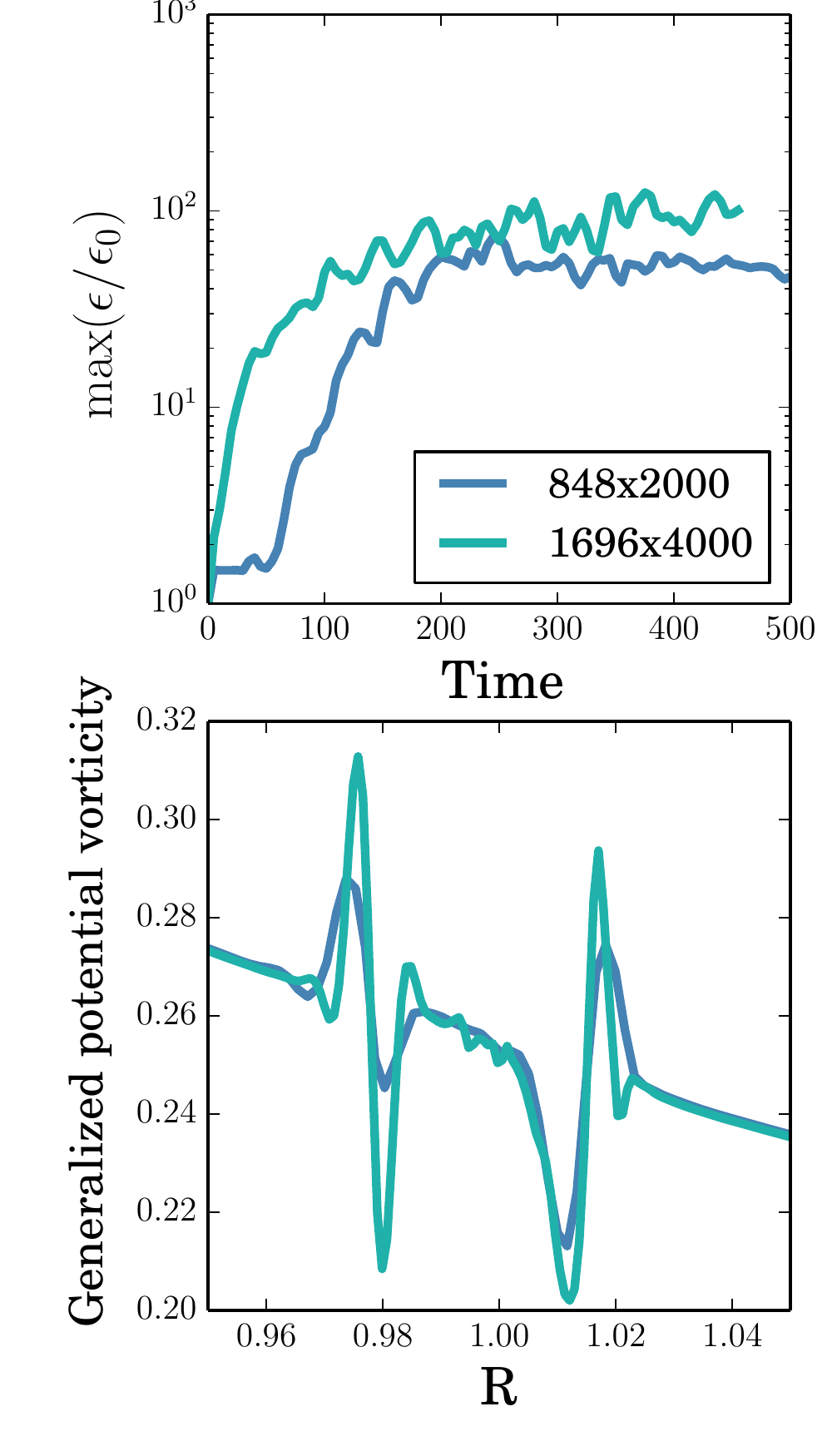}
\caption{{\it Left}: Relative pebble surface density perturbation at Time=50 for the high-resolution simulation  with ${\st=0.1}$ and $\epsilon=1$. {\it Right:}  
Maximum solid-to-gas ratio $\epsilon_{max}$ as a function of time (top), relative to the initial solid-to-gas-ratio $\epsilon_{0}$,  and  azimuthally averaged profile of the generalized potential vorticity (bottom)  for this run and for the fiducial, lower-resolution run. }

\label{fig:hr}
\end{figure}

\begin{figure}
\centering
\includegraphics[width=\columnwidth]{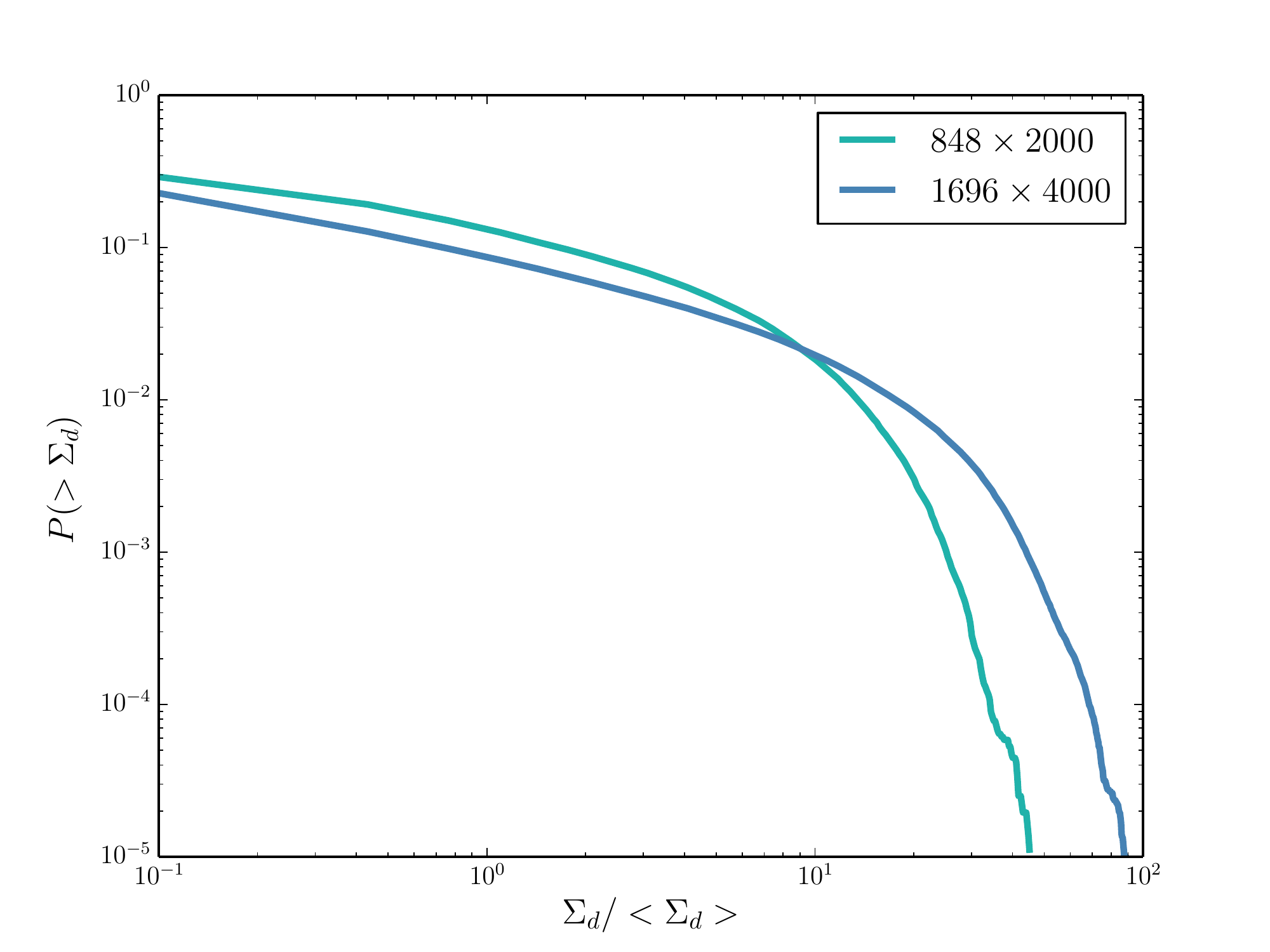}
\caption{ Cumulative pebble density distributions obtained from the moderate and high-resolution simulations. These have been obtained by counting the number of cells with pebble surface density above a certain value}

\label{fig:cdf}
\end{figure}

To check the convergence of our results, we performed an additional simulation for $\st=0.1$ and $\epsilon=1$ but with increased resolution ($N_R=1696,N_\phi=4000$).  A map of the pebble surface density  at Time=50  is shown in the  left panel of Fig. \ref{fig:hr}. Comparing  with that for lower resolution (upper left panel in Fig. \ref{fig:fiducial}), we see  multiple dusty eddies already forming at that time at the outer downstream separatrix  for the high resolution case. Thus, increasing the numerical resolution makes the instability be triggered earlier, and this is confirmed by inspecting $\epsilon_{max}$ as a function of time in the upper right panel of Fig. \ref{fig:hr}. This is due to the fact that gradients at the separatrix are better resolved (Chen \& Lin 2018), as can be observed in the lower right panel of Fig.  \ref{fig:hr} which displays the profiles of the generalized PV function at Time=50 and for both resolutions.  Moreover,  we see that the saturated value for $\epsilon_{max}$ tends to be higher by a factor of $\sim 2$ when doubling the resolution. This is also revealed by looking at the cumulative pebble density distributions in Fig. \ref{fig:cdf} and which have been obtained by counting the number of cells with pebble surface density above a certain value. Again, the instability that we describe here clearly shares various similarities with the classical streaming instability  for which previous studies have reported a lack of convergence when using a  fluid approach (Benitez-Llambay et al. 2019). This is because in the non-linear stage of the instability, the dynamics is dominated by crossing orbits such that the fluid approach is prone to fail because of shock formation  (Benitez-Llambay et al. 2019).   In the 2D inviscid simulations presented here, we also expect a lack of convergence due the small-scale clumps that emerge  at high numerical resolution. In the context of vortices forming at a planetary gap, McNally et al. (2019) have  also found that as the resolution is increased, the decreased numerical diffusion can give rise to small-scale scale vortices that can significantly impact the evolution of the system. It can be reasonably expected similar effects to be at work here, with  more and more clumps emerging at the planet separatrix as the resolution is increased.  Nevertheless,  we notice that numerical diffusion due to finite resolution  may not be unrealistic since  it could mimic the residual turbulence/viscosity in the dead zone.  Results of viscous simulations that include a small kinematic viscosity are presented in Appendix \ref{sec:visc}. In particular, these confirm that inviscid, moderate resolution calculations produce similar results as high resolution simulations that include a small level of kinematic viscosity. 

\subsection{Impact of Self-gravity}
\label{sec:sg}

An interesting question that emerges from the results of the simulations is whether the particle overdensities that are formed may collapse under the influence of self-gravity, giving rise eventually to the formation of planetesimals. While examining in details the effect of including self-gravity is beyond the scope of the paper, we nevertheless performed one run for $\epsilon=1$ and $\st=0.1$  and that includes the self-gravity of both the gas and pebble components. Because pebbles are considered as a pressureless fluid and since we  deal with an inviscid gas disc, we expect the disc of pebbles  to become gravitationally unstable when self-gravity is included.   In a more realistic disc, it is expected direct gravitational instability in the dust layer to require  dust densities which are orders of magnitude larger than roche density (Shi \& Chiang 2013),  due to effective pressure stabilization across the thin dust layer. Therefore,  it may not be easy to induce collapse in practice. Moreover,  other instabilities like the Kelvin-Helmholtz (Johansen et al. 2006) or streaming 
instabilities (Johansen \& Youdin 2007; Bai \& Stone 2010) may  prevent the dust from settling indefinitely.

 In order to bypass growth of the classical  gravitational instability, we therefoe added a pressure term in Eq. \ref{eq:dust} for the pebble velocity. This is equivalent to assuming a non-zero velocity dispersion $c$  and which we set such that $c^2=\delta c_s^2$ with $\delta=0.01$.  For this value of $\delta$, the Toomre parameter for the dust is ${\cal Q}_p=\kappa c/\pi G \Sigma_p\sim 2.5$ at the outer edge of the disc. However, we caution the reader that although the pebble and gas disc are gravitationally stable, the two-fluid might be unstable to Secular Gravitational Instability (SGI) due to the  gas-dust friction and the self-gravity of gas and dust (e.g. Takahashi \& Inutsuka 2014). For an inviscid disc, the instability criterion for the dusty disc is given by (Latter \& Rosca 2017):
\begin{equation}
{\cal Q}\lesssim \sqrt{1+\frac{\epsilon}{\delta}}
\end{equation}
or equivalently:
\begin{equation}
{\cal Q}_p\lesssim \frac{\sqrt{\epsilon+\delta}}{\epsilon}
\end{equation}

\begin{figure}
\centering
\includegraphics[width=\columnwidth]{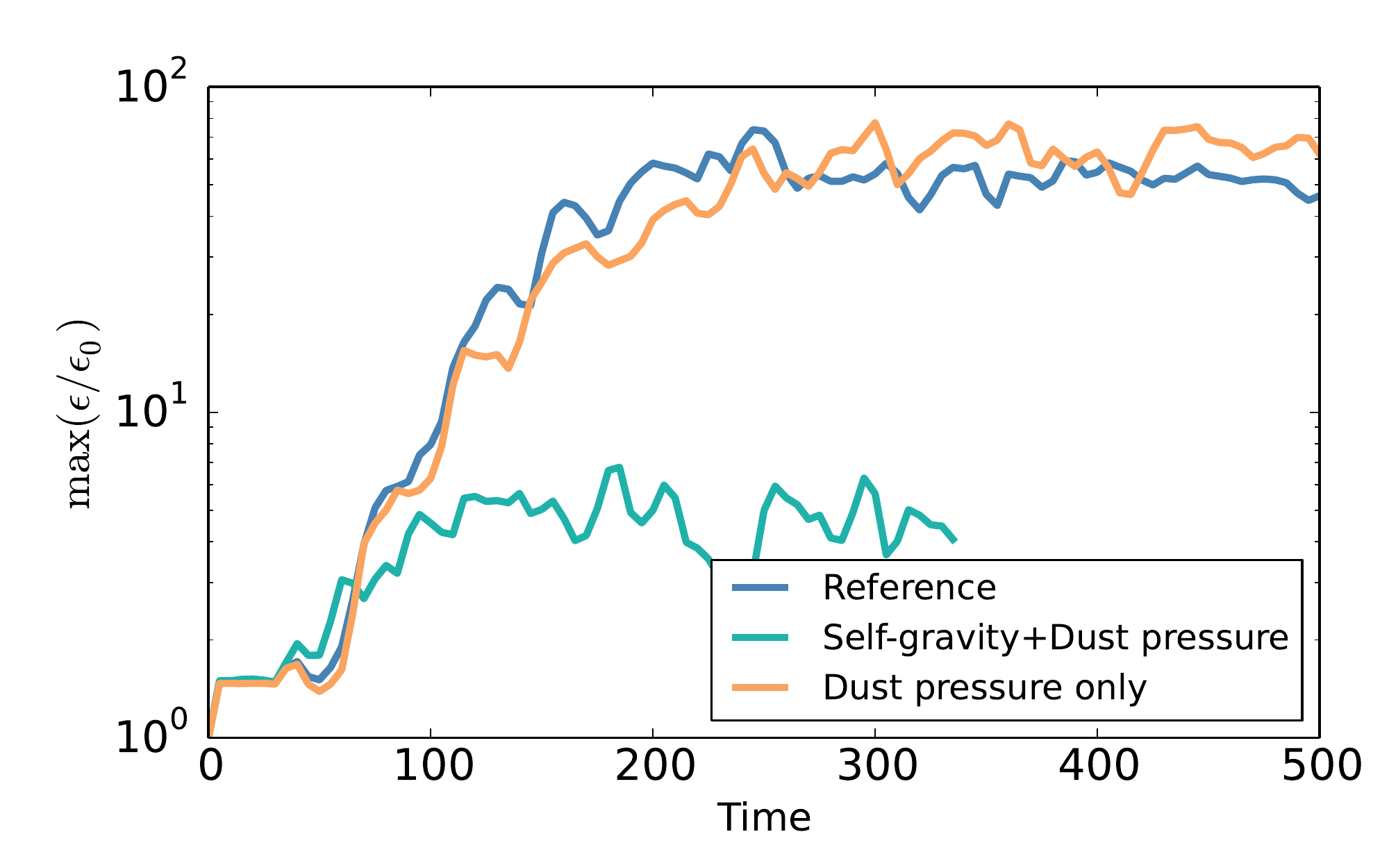}
\caption{ 
Maximum solid-to-gas ratio $\epsilon_{max}$ as a function of time (top), relative to the initial solid-to-gas-ratio $\epsilon_{0}$, for runs that include or discard the effect of disc self-gravity.  }
\label{fig:ratiosg}
\end{figure}

For $\delta=0.01$ and $\epsilon=1$, this gives ${\cal Q}\lesssim 10$ and ${\cal Q}_p\lesssim 1$ so that we do not expect the SGI to be present in our simulations. \\
 In Fig \ref{fig:ratiosg} is presented the time evolution of $\epsilon_{max}$ for the run including self-gravity. Compared to the case where self-gravity is discarded, we see  that particle concentrations are much more weaker, with a value of $\epsilon_{max}\sim 5$ at saturation.  We note that this does not occur because of the added extra pressure in the dust component, since a run performed without self-gravity and including this pressure term resulted in a saturated value for $\epsilon_{max}$ similar to that inferred from the fiducial calculation. This could rather possibly occur because in presence of self-gravity, i) vortices are significantly weakened (Zhu \& Baruteau 2016)  resulting in a reduced efficiency of particle trapping  and ii)   merging of dusty vortices can be avoided because of mutual horseshoe U-turns (Lin \& Papaloizou 2011) or scattering.\\

\begin{figure}
\centering
\includegraphics[width=\columnwidth]{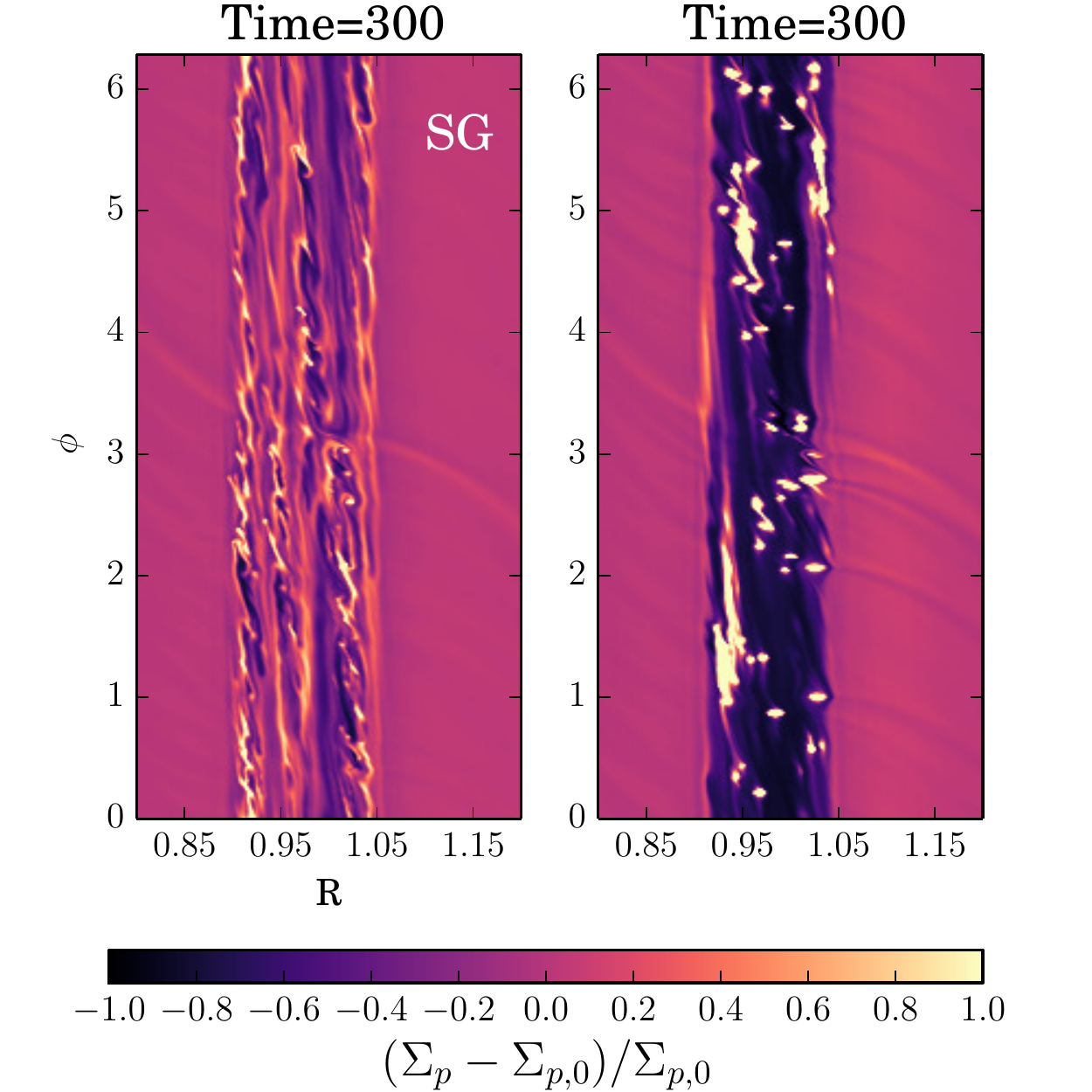}
\includegraphics[width=\columnwidth]{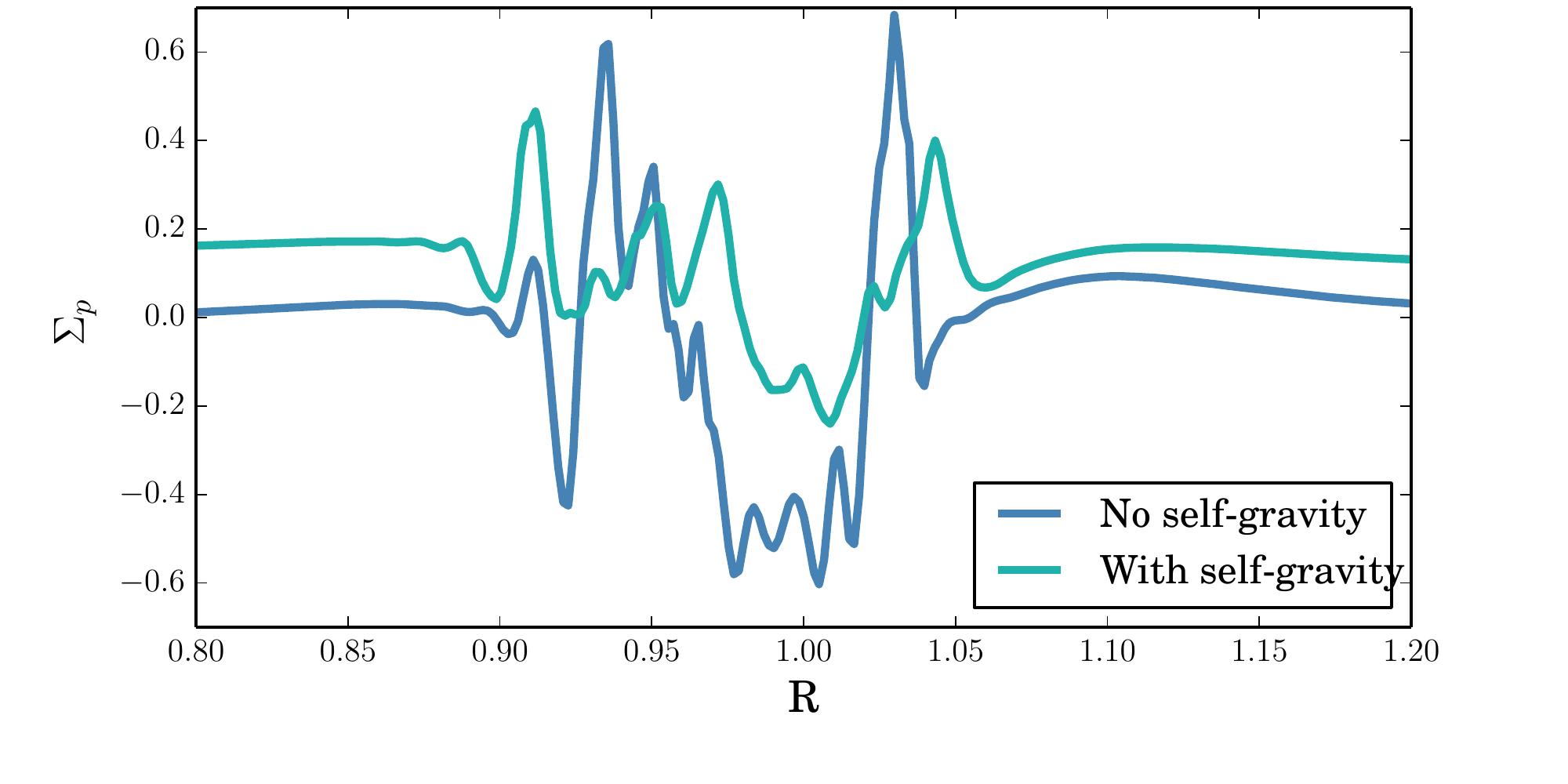}
\caption{{\it Top}: For ${\st=0.1}$ and $\epsilon=1$, comparison of the relative pebble surface density perturbation at Time=300 between the simulation  with self-gravity (left) and  the simulation without self-gravity included (right). In the simulation that includes self-gravity, the smoothing parameter was set to $b_d=0.1b$. {\it Bottom:} Corresponding pebble surface density profiles at the same time.}
\label{fig:sgdust}
\end{figure}

Fig. \ref{fig:sgdust} shows the pebble surface density at Time=300, for runs with (left panel) and without (right panel) self-gravity. Including self-gravity tends to produce narrow rings of particles and spiral structures rather than dense filaments, which suggests that self-gravity tends to counterbalance the effect of particle feedback. We note in passing  that such multiple dust rings and vortices forming at either side of the planet gap have also been observed in previous studies of Super-Earths in low-viscosity, non self-gravitating discs. (Dong et al. 2017, 2018). The corresponding surface density profiles in the bottom panels also reveals a trend for the gap to be shallower in presence of self-gravity. Compared to the non self-gravitating case, this implies a higher  effective viscosity when  self-gravity is considered, which could  be due to additional gravitational stresses. The fact that  more vortices are produced in that case also contributes to a shallower gap by providing more transport within the gap region.   

\section{Implication for planet migration}
\label{sec:planet}
\begin{figure}
\centering
\includegraphics[width=\columnwidth]{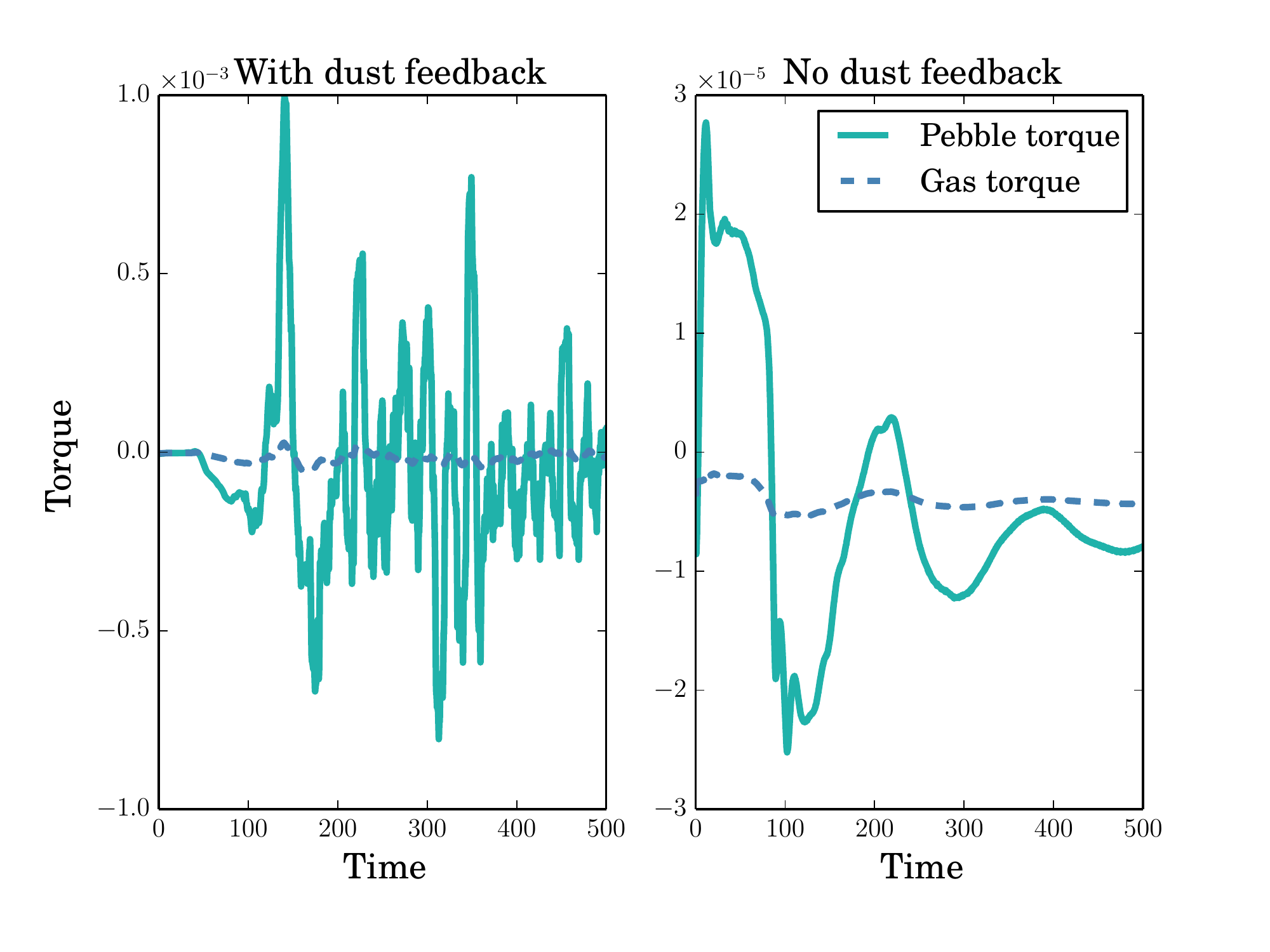}
\caption{ {\it Left:} gas and pebble torques as a function of time for the run with  ${\st=0.1}$ and $\epsilon=1$. {\it Right:} same but in the case where the impact of dust feedback is not considered. We note that here, the planet evolves on a fixed orbit.}
\label{fig:torque}
\end{figure}

\begin{figure}
\centering
\includegraphics[width=\columnwidth]{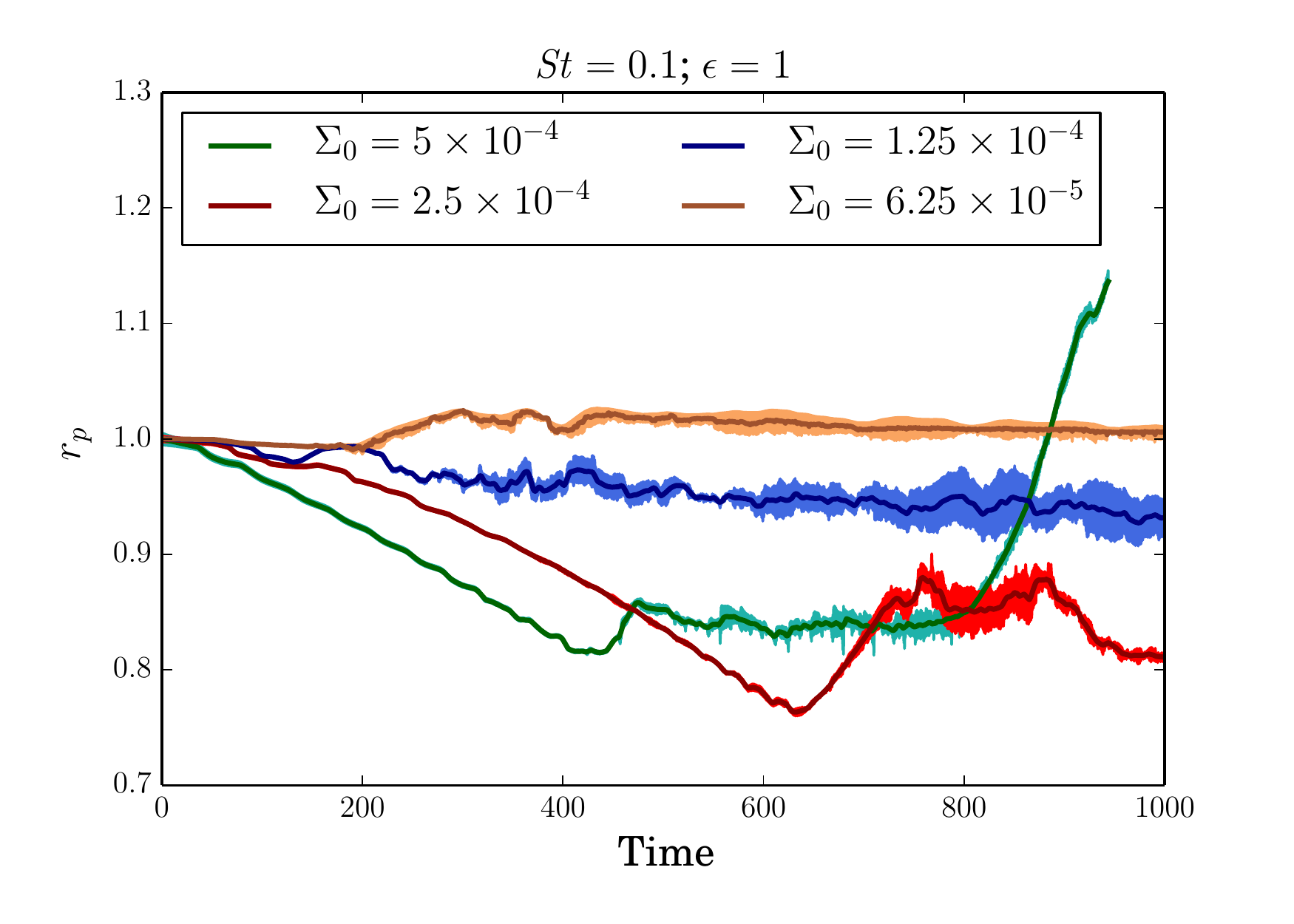}
\caption{ Planet orbital distance as a function of time for  runs with  ${\st=0.1}$ and $\epsilon=1$ and in which the planet is allowed to migrate. The filled area is enclosed between the two curves corresponding to the time evolution of the planet pericenter and apocenter. }
\label{fig:rp-migrate}
\end{figure}

\begin{figure}
\centering
\includegraphics[width=\columnwidth]{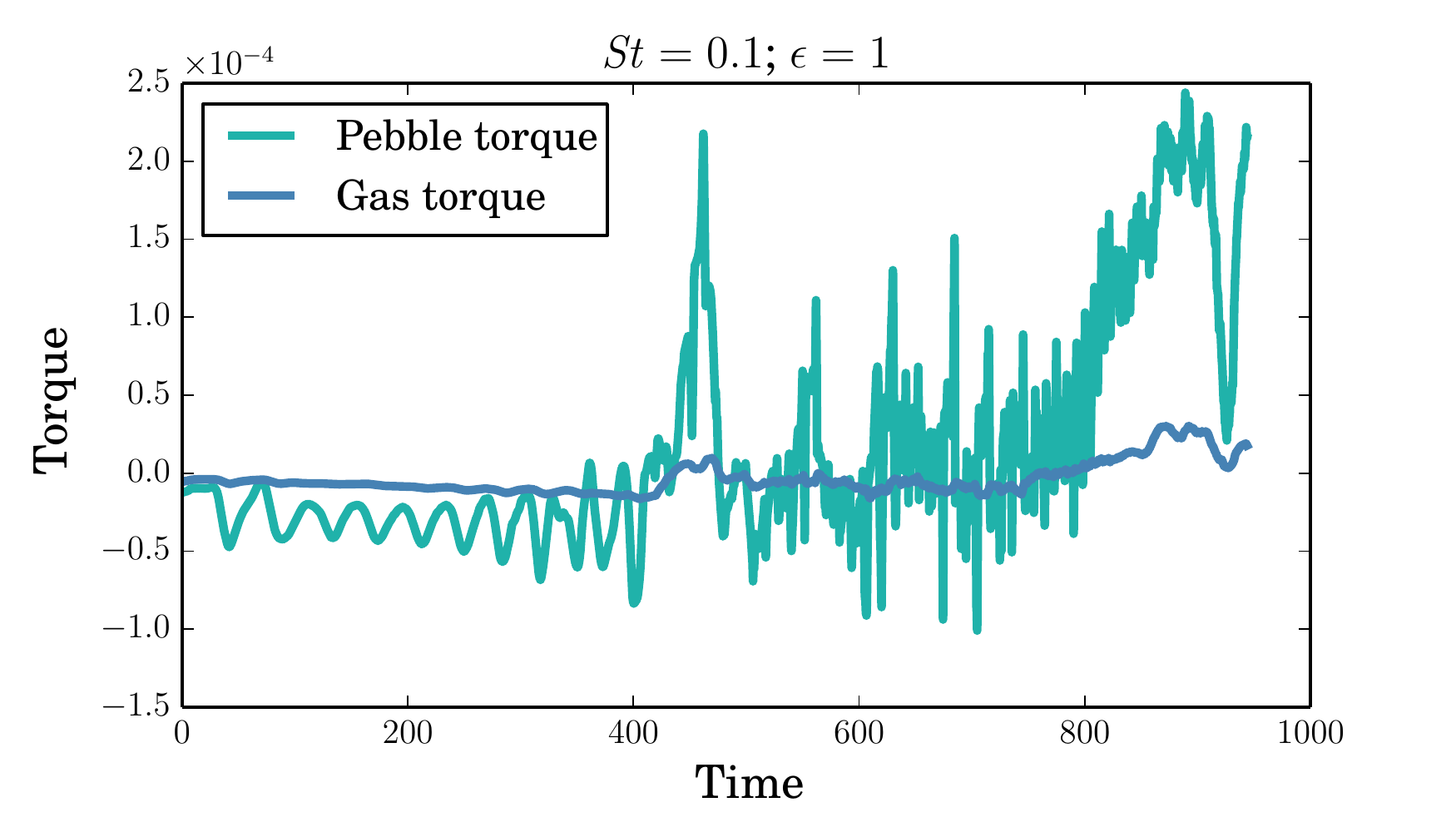}
\caption{Gas and pebble torques as a function of time for the run where the planet is allowed to migrate and with $\Sigma_0=5\times 10^{-4}$.}
\label{fig:torque-migrate}
\end{figure}

\begin{figure*}
\centering
\includegraphics[width=\textwidth]{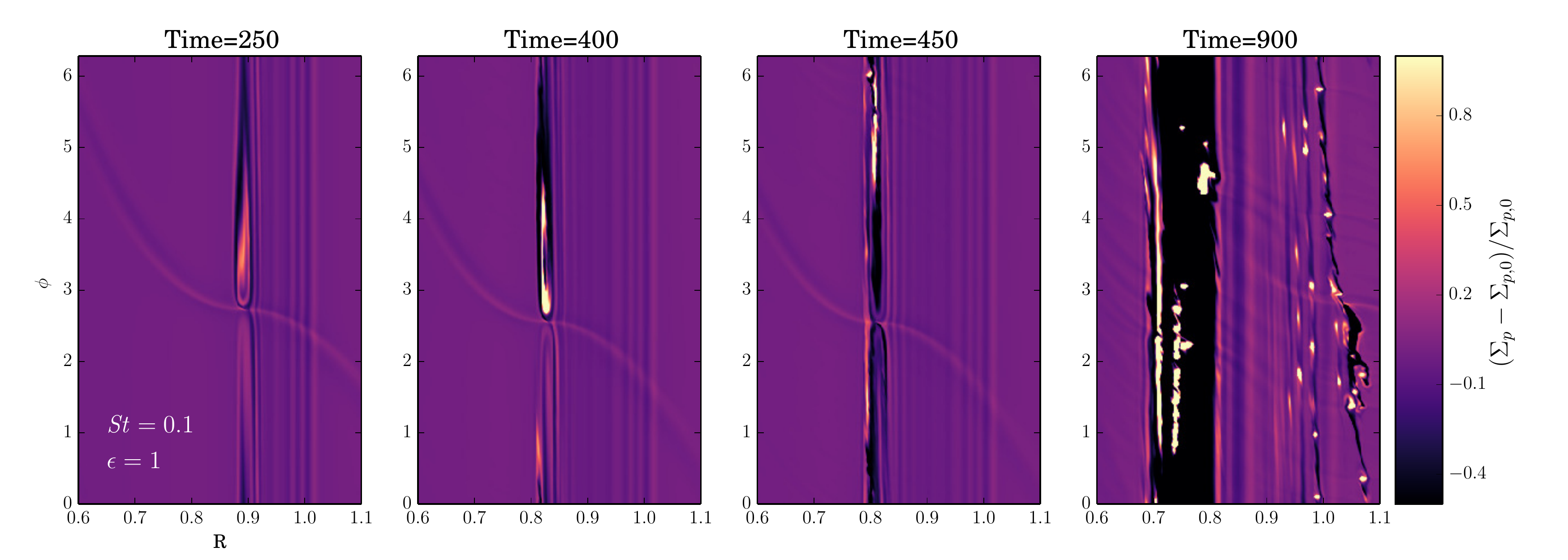}
\caption{ Evolution of the relative pebble surface density perturbation for the run with ${\st=0.1}$ and $\epsilon=1$ but in which the planet is allowed to migrate}
\label{fig:migrate-dust}
\end{figure*}

\begin{figure}
\centering
\includegraphics[width=\columnwidth]{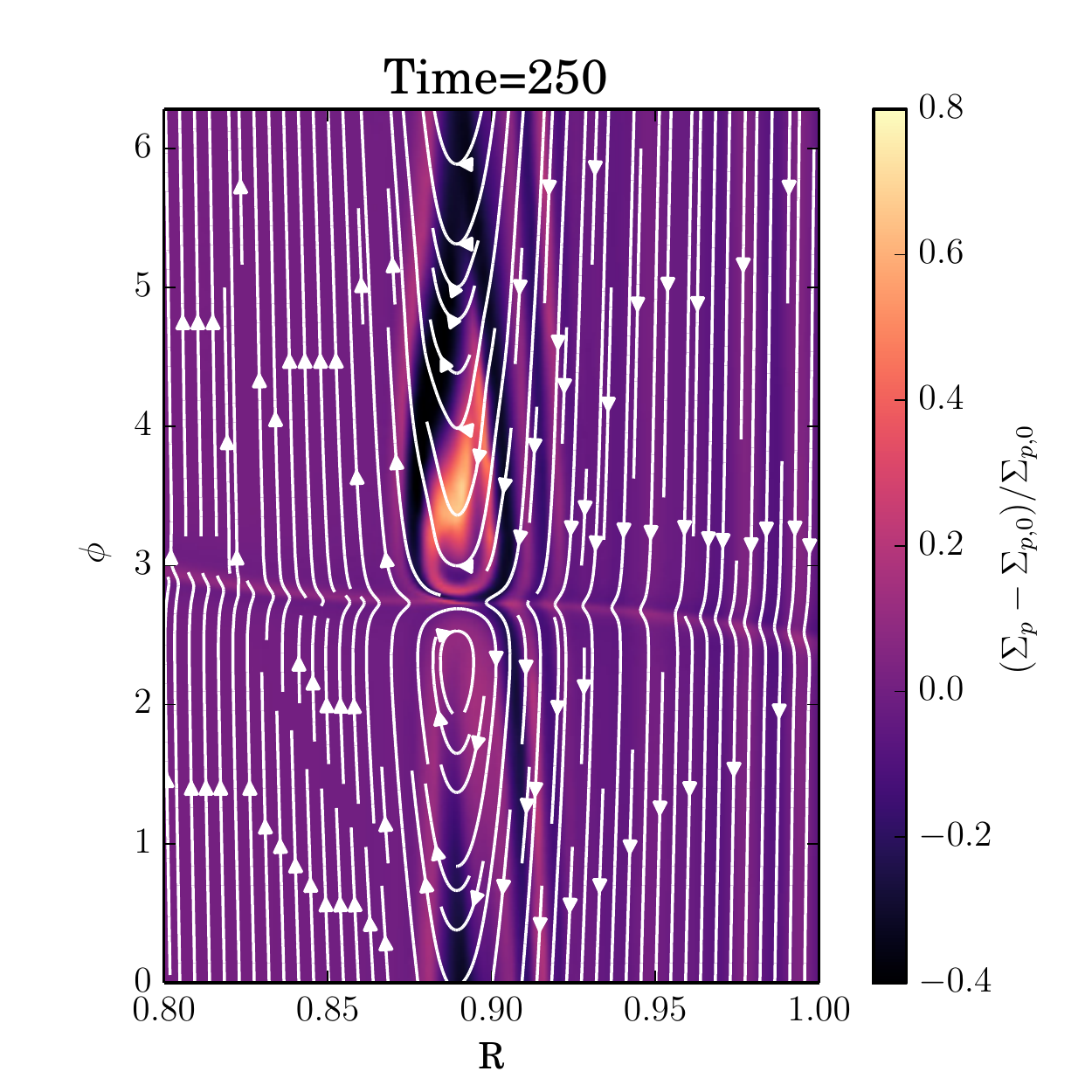}
\caption{Streamlines and relative surface density perturbation of pebbles at Time=250 for  the simulation in which the planet is allowed to migrate.}
\label{fig:streamlines}
\end{figure}

In this section, we discuss the implication of our results in the context of planet migration. For inviscid discs, previous studies have demonstrated the important role of the formation of vortices on the evolution of planets that are massive enough to induce a strong change in the gas surface density profile (Li et al. 2009; Yu et al. 2010; McNally et al. 2019). Chen \& Lin (2018) have shown that a low-mass planet embedded in a dusty disc can also undergo large torque oscillations in the case of strongly coupled dust with $\st=10^{-3}$, either due to the formation of a blob in the potential vorticity for dust-to-gas ratios $\epsilon \sim 0.5$, or to vortex instability for higher metallicities.\\
In  the left panel of  Fig. \ref{fig:torque} we show the gas and pebbles torques exerted on the planet for the fiducial simulation with $\st=0.1$ and $\epsilon=1$.  On average, the gas torque remains negative,  with a value which is close to that corresponding to the Lindblad torque $\Gamma_L$ and which is given by {(Paardekooper et al. 2010)}:
\begin{equation}
\Gamma_L=(-0.85+\sigma)(q/{\tilde h})^2\Sigma a_p^4 \Omega_k^2
\end{equation}
where all quantities are evaluated at the location of the planet and where $\tilde h=\tilde H/R$, with $\tilde H$ given by  Eq. \ref{eq:heff}.
This is not surprising since in an inviscid disc,  the corotation torque is expected to cancel after a few horsehoe libration timescales due to saturation process ( Balmforth \& Korycansky 2001).  This saturation process requires typically $\sim 5$ libration timescales (e.g. Kley \& Nelson 2012) and is clearly visible in the right panel of Fig. \ref{fig:torque} where we show the evolution of the torques in the case where the effect of dust feedback is not taken into account.  Turning back to the left panel of Fig. \ref{fig:torque}, we observe, over long timescales ,  a slight tendency for the amplitude of the gas torque to decrease with time, due the accumulation of gas material inside of the planet orbit and which exerts a positive the torque on the planet (see Time=500 panel in the middle row of Fig. \ref{fig:fiducial}).  Compared to the gas torque, the pebble torque exhibits a much more erratic behaviour. Before the RWI sets in, we see that the pebble torque is negative because  i) the rear side of the downstream separatrix which exerts a negative torque on the planet tends to be overdense compared to the rest of the disc (see Time=50 panel in the top row of Fig. \ref{fig:fiducial}) and ii) part of the corotation region located ahead of the planet is slightly depleted (see Time=100 panel in the top row of Fig. \ref{fig:fiducial}).  At later times, the pebble torque undergoes large oscillations due to vortex formation, although it remains negative on average. This occurs because, as mentionned in Sect. \ref{sec:st0v1}, the inner edge of the pebble gap continuously moves away from the planet, resulting in a decrease of the positive torque exerted by the inner disc. From the evolution of the gas and pebble torques, we would therefore naively expect the planet migration to be somewhat chaotic due to the interaction of the planet with dusty eddies.   Additional simulations in which the planet is allowed to migrate as soon as it is introduced  demonstrate that in fact this is not necessarily the case.  This is illustrated in Fig. \ref{fig:rp-migrate} where we show the planet orbital distance $r_p$ (bottom panel) as a function of time for runs where the planet is allowed to migrates and that differ by the surface density of the gas disc. For $\Sigma_0=5\times 10^{-4}$,  the planet initially undergoes a rapid inward migration episode before migration stops at Time=400 and reverses at Time=800. Examination of the torques in Fig. \ref{fig:torque-migrate} clearly  reveals that migration is mainly driven by the pebbles and we  display in Fig.  \ref{fig:migrate-dust} maps of the pebble surface density at different times.  Estimating the planet drift rate $\dot a_p$ from Fig. \ref{fig:torque-migrate} together with the pebble drift velocity $V_{rel}$   results in a relative drift speed beween the planet and the dust of 
$v_{rel}\sim 0.3 {\dot a}_p$, corresponding to a  relative drift timescale across the horseshoe region of $t_{drift}=x_s/v_{rel}\sim 95$ orbits. This is shorter than horseshoe libration timescale $t_{lib}\sim 8\pi a_p/3\Omega x_s\sim 120$ orbits,  
which implies that there is a small inward flux of pebbles in the planet corotation region. Because the migration timescale is short enough for the planet to cross the horseshoe region over one horseshoe libration time, the corotation region becomes significantly distorted, particularly ahead of the planet (see Time=250 panel). However, we remark that the teardrop shape that is seen does not result from dynamical torques acting on the planet (McNally et al. 2018). This can be  demonstrated by inspecting  the (pebble) streamlines which are plotted at Time=400 in Fig. \ref{fig:streamlines}. Here, the trapped tadpole region is located behind of the planet and is therefore not related to the high density region forming ahead of the planet.  Nevertheless, such a teardrop-like region is expected to exert a potentially strong positive torque on the planet.  As revealed by the Time=400 panel, this positive torque increases as the planet migrates because the pebble surface density within this region becomes higher,  although this region does not correspond to a pressure bump. We checked that this results from an increase in the amount of pebble material in this region,  possibly coming from pebbles that are scattered  at the inner downstream separatrix and which become subsequently trapped ahead of the planet due to migration.  This occurs up to a point in time where the positive torque exerted by this region becomes high enough to stop migration.  Subsequent evolution consists in the planet evolving on a constant trajectory with almost fixed semimajor axis for $\sim 400$ orbits, which is a sufficient period of time to trigger vortex instability at the inner edge of  the planet separatrix.  Vortex formation  induces excitation of the planet eccentricity and subsequently leads to the rapid outward migration of the planet (McNally et al. 2019). Interestingly, this episode of outward migration appears to be sustained because i) the flow of pebbles across the orbit of the planet has a positive feedback on migration  because pebbles that are scattered inwards exert a positive torque on the planet and ii) there is an underdense librating region at the trailing side of the planet which  has a negative feedback on migration. This suggests that under certain conditions that we will examine in more details in a future paper,  a low-mass planet migrating in a dusty disc might be subject to strong dynamical torques (Paardekooper 2014; Pierens 2015) from the solid component. Interestingly, the Time=900 panel shows that although the planet has migrated outward to $r_p\sim 1.05$,   elongated structures typical of the instability that we described in this paper have been formed and evolve  in a common gap 
centered at $0.7<R<0.8$.  This could have important observation implications because it suggests that planet-induced dust gaps need not contain planets.  Multiple rings and isolated vortices can also be observed in the region $0.8<R<1.05$, which is consistent with the results of McNally et al. (2019) who found that multiple rings and vortices can persist long after the planet has passed. \\
 This episod of fast ouward migration is not observed to occur in simulations adopting a lower surface density, although these confirm that the migration of the planet tends to delay the onset of the instability. Reduction of the disc mass by a factor of 2 resulted in the planet migration to be almost stalled as the instability is rapidly triggered and propagates away from the planet orbit (see Sect. \ref{sec:st0v1}).  Generally speaking,  we expect the evolution outcome to depend significantly on the value for the pebble flux across the corotation region and therefore on the value for the disc surface density. In the case of an inward migrating planet with a drift speed much smaller that the pebble radial velocity, pebbles can catch up with the planet and can exert a positive torque when they are scattered inward by the planet. Such an effect would be even more pronounced for a planet migrating outwards. We will examine in more details  the effect of varying the pebble mass flux in a future paper.

\section{Conclusions}
We have presented a series of two-fluid hydrodynamical simulations of low-mass planets embedded in inviscid protoplanetary discs and interacting with inward drifting pebbles. The discs that we considered have relatively high solid-to-gas ratio $\epsilon \ge 0.5$, which might represent the typical 
conditions expected in the innermost parts of the disc as a result of dust feedback or  magneto-thermal winds. Our main conclusion is that  dusty  vortices can develop through the Rossby Wave Instability (RWI) at the planet separatrix between librating and circulating streamlines, where  
extrema in the  generalized potential vorticity profile can exist.  Depending on the value for the Stokes number, different evolution outcomes are obtained.  In the case with $\st=0.01$, a single vortex that captures pebbles is formed at the planet separatrix. Accumulation of solids inside the vortex proceeds  until the maximum dust-to-gas ratio saturates at $\epsilon_{max}\sim 50$ once the amplitude of the Rossby number at vortex center starts to decrease. For pebbles with Stokes numbers $\st= 0.1, 0.5$, however,   multiple dusty vortices are formed at the planet separatrix. These dusty clumps have enhanced dust-to-gas ratio such that they orbit slower and suffer  reduced radial drift compared to rest of solid particles. As a consequence,  pebbles can catch up with these dusty clumps,  which can  lead to a further increase in the solid-to-gas ratio. For $\st= 0.1$ increase in the solid-to-gas ratio of approximately two orders of magnitude is obtained, whereas we find an enhancement in the solid-to-gas ratio of $\sim 10^3$  for pebbles with $\st \sim 0.5$ that are marginally coupled to the gas.  For this range of Stokes numbers, the instability that is triggered at the separatrix shares similarities with the classical streaming instability. In particular, we find that the non-linear stage of the instability consists of multiple azimuthally extended structures (or filaments) orbiting in a gas-depleted disc. The main difference between these two instabilities  is that the radial wavelength of the instability developing at the planet separatrix is typically a fraction of the pressure scale height  $H_g$ whereas the one corresponding to the streaming instability is $~10^{-4}-10^{-3}$ $H_g$.  \\
We examined the issue of whether the azimuthally structures that are formed through this process may collapse to form planetesimals by considering the effect of self-gravity. We showed that including self-gravity leads  in fact to lower solid-to-gas ratios enhancements, possibly because self-gravity is known to weaken vortices and/or prevent vortex merging via mutual horseshoe U-turns. In that case, elongated structures are not produced but we rather observe narrow rings of solids forming at either side of the planet orbit, and with a clear trend for dust gaps to be shallower. This would suggest  that dust gaps are shallower in younger, more massive discs.  The precise impact of   including self-gravity has however to be investigated in more details, and this will be the subject of a future study. \\
We also found that the dusty vortices forming at the separatrix of a low-mass planet can play a significant role on its orbital evolution. For a planet held on a fixed orbit, the torques induced by the pebbles indeed exhibit large amplitude oscillations as a result of the instability.  In the case where the planet is allowed to migrate in a disc with mass corresponding to the MMSN, however, we find that vortex formation is delayed as a consequence of the planet drift, which is rapid enough to significantly distort the corotation region.   A  high density of solid material  instead forms ahead of the planet, which exerts a positive torque and causes migration to be halted. This allows the vortex instability to set in, while the planet is observed to undergo an episode of rapid outward migration possibly due to dynamical torques.  Interestingly, this occurs while a dust  gap is forming inside the planet orbit. A related consequence to this result is that planetary dust gaps may not necessarily contain planets if these migrated away. In that case, a dust gap  would rather  form at the location where the planet migration  reversed and may still contain dust clumps as vestiges of disc-planet interactions.  For much lower disc masses, however, we find an almost stalled migration due to the rapid onset of the instability at planet separatrix.   \\
 An important caveat of this work is the use of a fluid approach to study the non-linear stage of the instability, which corresponds to a regime that can be dominated by crossing orbits. In this regime, the fluid approach may not be applicable due to shock formation and it would therefore be valuable to confirm or refute our findings using a more appropriate method based on  Lagrangian particles (Bai \& Stone 2010).   
Moreover, the resolution that we adopted in our simulations is not high enough to examine potentially important effects that might occur within the vortex core.  For example, it has been shown that a vortex can be destroyed by a dynamical instability due to  the effect of dust feedback and that can occur once the solid density within its core becomes comparable to the gas density (Fu et al. 2014; Crnkovic-Rubsamen et al; Raettig et al. 2015). In that case, such a process could play an important role on the non-linear outcome of the vortex instability presented here.  On the other hand, recent work (Lyra et al. 2018) suggests that the pebble backreaction  does not destroy  vortices when adopting a 3D setup, because pebbles alter the vortex structure only in the disc midplane in that case. Clearly, more work is required, eventually using 3D simulations, to evaluate the consequences on planetesimal formation and planet migration of forming dusty vortices at the separatrix of a low-mass planet. Moving to a 3D setup  would also allow to evaluate the consequences of a solid-to-gas ratio that varies with height as a consequence of dust settling.   

\section*{Acknowledgments}
Min-Kai Lin is supported by the Theoretical Institute for Advanced Research in Astrophysics (TIARA) based in the Academia Sinica Institute for Astronomy and Astrophysics (ASIAA) and the Ministry of Science and Education (MOST) grant 107-2112-M001-043-MY3.Computer time for this study was provided by the computing facilities MCIA (M\'esocentre de Calcul Intensif Aquitain) of the Universite de Bordeaux and by HPC resources of Cines under the allocation A0050406957 made by GENCI (Grand Equipement National de Calcul Intensif).

\appendix

\section{Radial drift problem}
\label{sec:test}
\begin{figure}
\centering
\includegraphics[width=\columnwidth]{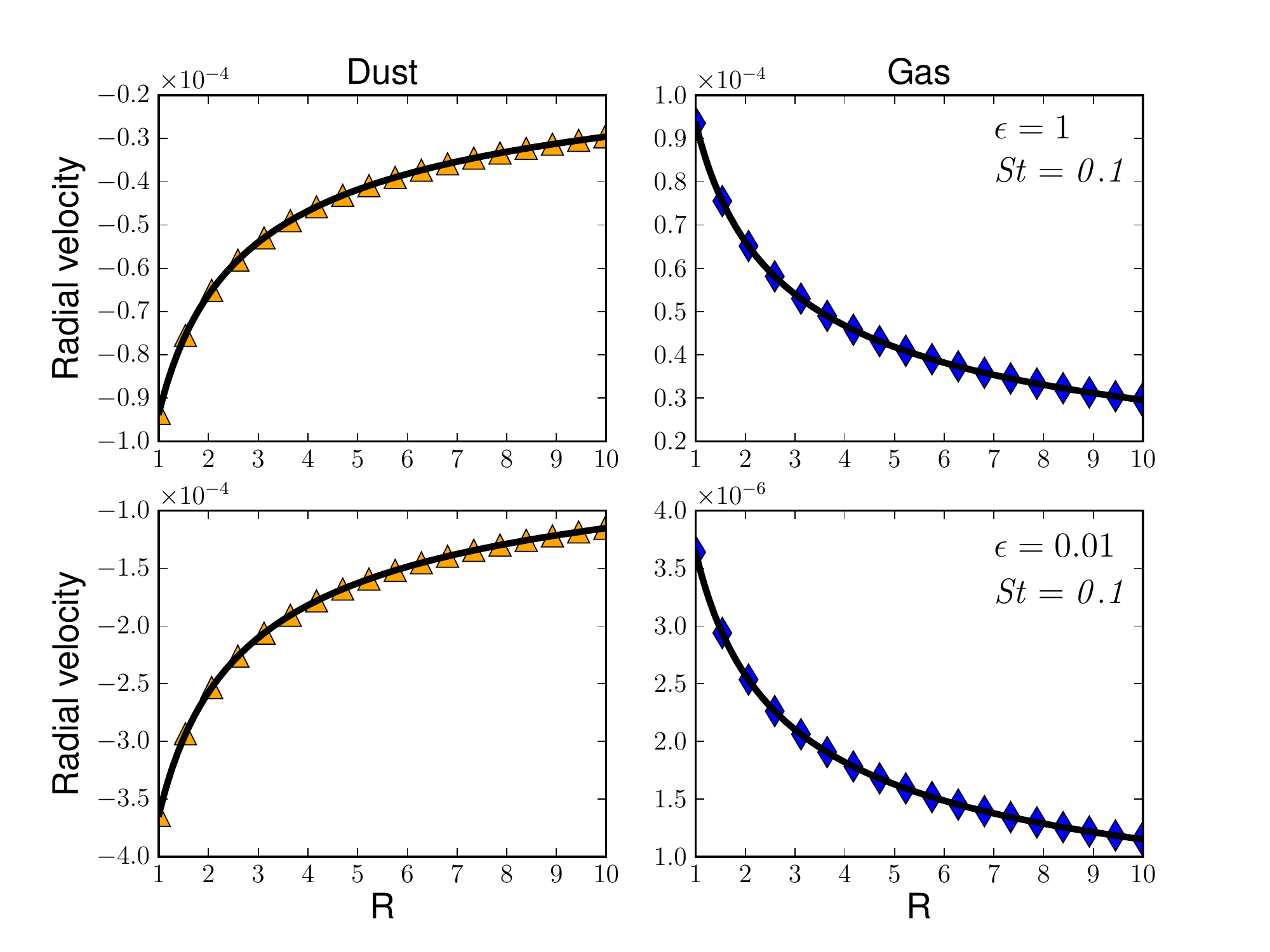}
\caption{Radial velocity of the dust and gas components for runs aimed at testing the method to solve the momentum transfer between dust and gas through the radial drift problem. Analytical solutions given by Eqs. \ref{eq:vrd} and \ref{eq:vrg} are plotted with solid line while numerical simulations are displayed with colored symbols. }
\label{fig:test}
\end{figure}

Here, we show that our implementation of the method to solve the momentum transfer between the dust and gas accurately reproduces the steady-state solutions given by Eq. \ref{eq:vrd}-\ref{eq:vtg}. For this test, the disk extends from $R_{in}=1$ to $R_{out}=10$ and we use $N_R=512$ uniformly distributed grid cells.  The initial conditions are similar to those presented in the main text, namely the initial gas surface density profile has power-law index $s=-1/2$ and the aspect ratio is constant with $h=0.05$. Again, we use damping zone near the boundaries where the velocities and densities are relaxed to their initial value. Here, the buffer region extends from $R=1.0$ to $R=1.31$ at the inner boundary while it is located in the range $7.65<R<10$ at the outer boundary. 
For this test, we considered  pebbles with Stokes number $\st=0.1$ (corresponding to the value adopted in the reference run presented in the main text) and two values for the dust-to-gas ratio namely $\epsilon=0.01,\; 1$.  For this radial drift problem, we compare in Fig. \ref{fig:test}  the dust and gas radial velocities obtained  from our calculation with the corresponding analytical solutions given by Eqs. \ref{eq:vrd} and \ref{eq:vrg}.  Very good agreement is obtained in each case, which validates the numerical method used in this work.

\section{Results for simulations with non-zero kinematic viscosity}
\label{sec:visc}
\begin{figure}
\centering
\includegraphics[width=\columnwidth]{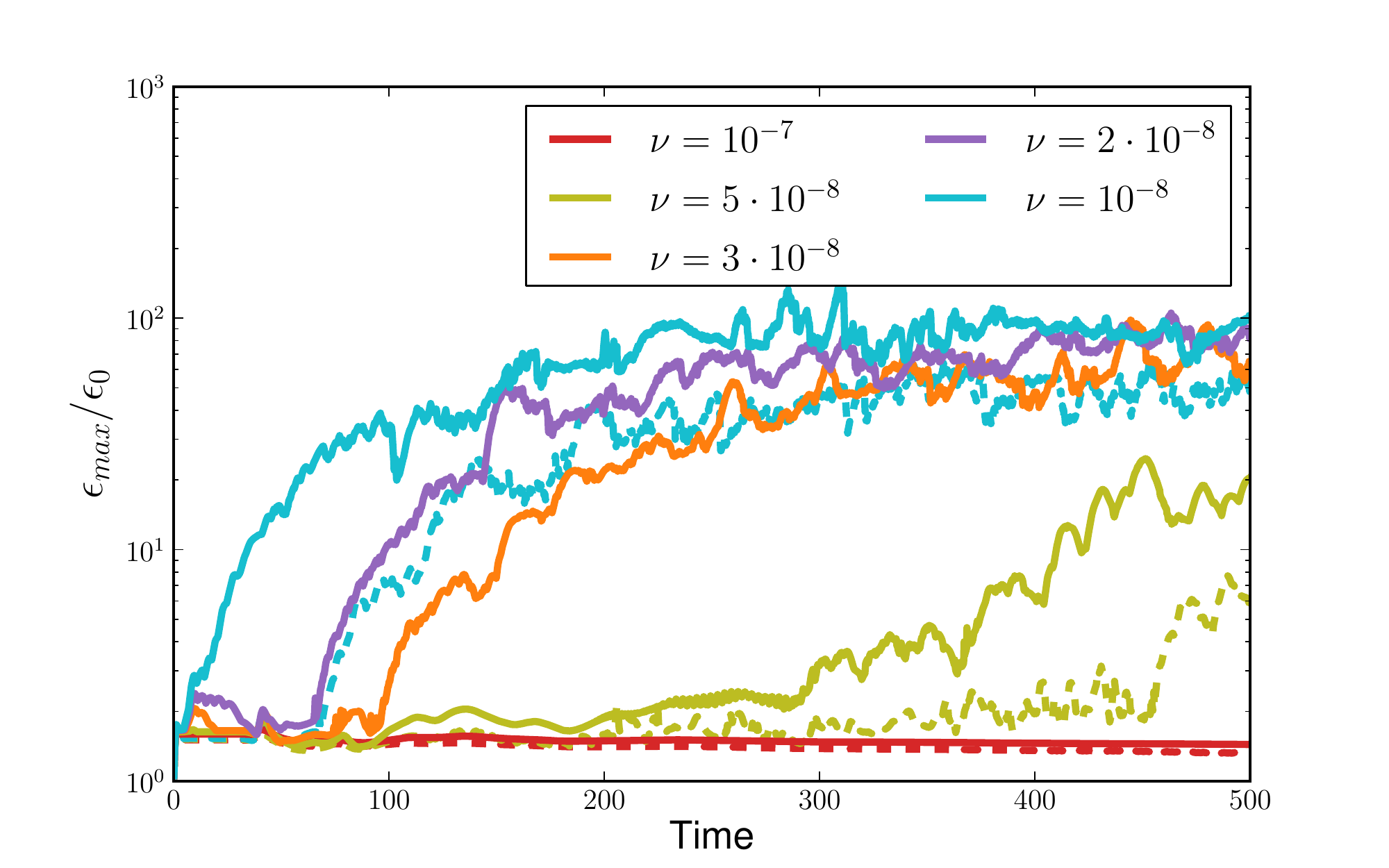}
\caption{Maximum solid-to-gas ratio $\epsilon_{max}$ as a function of time, relative to the  initial solid-to-gas ratio $\epsilon_0=1$, for different levels of kinematic viscosity. Dashed lines correspond to moderate resolution runs with $(N_R,N_\phi)=(848,2000)$ whereas solid lines are for high resolution simulations with $(N_R,N_\phi)=(1696,4000)$.  }
\label{fig:viscous_runs}
\end{figure}

 In the inviscid limit that we considered in this paper, the numerical dissipation may significantly affect our results, such that these could be dependent on the adopted resolution.  Here, we present the results of additional simulations that employ a non-zero value for the kinematic viscosity. The aim is to study the dependence of the results when changing the dissipation in a controlled manner.  For this set of runs, we adopted parameters corresponding to the reference simulation with $\st=0.1$ and $\epsilon_0=1$, but  varied the viscosity  in the range $10^{-7}\le\nu\le
  10^{-8}$  at two grid resolutions with $(N_R,N_\phi)=(848,2000)$ and  $(N_R,N_\phi)=(1696,4000)$. The time evolution of $\epsilon_{max}$ for these runs is presented in Fig. \ref{fig:viscous_runs}.  At both grid resolutions, growth of $\epsilon_{max}$ is not observed for $\nu=10^{-7}$, which is equivalent to a viscous stress parameter $\alpha=4\times 10^{-5}$.  This implies that the instability really requires a nearly laminar disc, similarly to the classical streaming instability which needs typically $\alpha \lesssim 10^{-5}$ (e.g. Auffinger \& Laibe 2017).  For kinematic viscosities with $\nu\le 5\times10^{-8}$, however,  the instability is triggered at both grid resolutions, with an increased growth rate and higher saturated value of $\epsilon_{max}$ as the viscosity is decreased. For clarity, we do not show the evolution of $\epsilon_{max}$ for the inviscid simulations that we presented in the main text but comparison with  the results for $\nu=10^{-8}$ clearly reveals that the linear and non linear stages of the instability proceed very similarly in both cases, and for the two grid resolutions that we adopted.  Thus, we estimate the numerical visosity to be close to $\sim 10^{-8}$. More importantly, we see that the evolution of $\epsilon_{max}$ in the high resolution run with 
  $\nu=2-3\times 10^{-8}$ is close to that inferred from the moderate resolution, inviscid calculation. This suggests that the inviscid moderate resolution runs that we presented in the paper are equivalent to high-resolution simulations with a low level of kinematic viscosity  $\nu=2-3\times 10^{-8}$.

\end{document}